\documentclass[%
 reprint,
superscriptaddress,
 amsmath,amssymb,
 aps,
]{revtex4-1}
\usepackage[T1]{fontenc}

\usepackage{alemfonts}
\usepackage[utf8]{inputenc}

\usepackage{amsfonts}
\usepackage{amssymb}
\usepackage{mathrsfs}
\usepackage{amsthm}
\usepackage{float}
\usepackage{graphicx}
\usepackage{dcolumn}
\usepackage{bm}
\usepackage{xcolor}
\usepackage{amsmath}

\usepackage{accents}

\newcommand{\vardbtilde}[1]{\tilde{\raisebox{0pt}[0.85\height]{$\tilde{#1}$}}}

\long\def \beq#1\eeq {\begin{align} #1 \end{align}}
\long\def \beaq#1\eeaq {\begin{align}\begin{aligned} #1 \end{aligned}\end{align}}
\long\def \bes#1\ees {\begin{align}\begin{split} #1 \end{split} \end{align}}
\long\def \bea#1\eea {\begin{eqnarray} #1 \end{eqnarray}}
\long\def \bse[#1]#2\ese {\begin{subequations}\label{#1}\begin{align} #2 \end{align}\end{subequations}}

\newcommand{\mv}[1]{\langle #1\rangle}

\long\def\dm[#1]{\!\operatorname{d\mu}\left(#1\right)}

\newcommand{\RS}{{\rm \scriptscriptstyle RS}}
\newtheorem{Remark}{Remark}
\newtheorem{Theorem}{Theorem}

\newtheorem{Definition}{Definition}

\graphicspath{{./Figures/}}

\begin{document}

\preprint{APS/123-QED}

\title{Neural networks with redundant representation:\\ detecting the undetectable}

\author{Elena Agliari}
 \email{agliari@mat.uniroma1.it}
\affiliation{%
Dipartimento di Matematica ``Guido Castelnuovo'', Sapienza Universit\`a di Roma, Roma, Italy
}%

\author{Francesco Alemanno}%
\affiliation{%
Dipartimento di Matematica e Fisica ``Ennio De Giorgi'', Universit\`a del Salento, Lecce, Italy
}%
\affiliation{C.N.R. Nanotec, Lecce, Italy}%

\author{Adriano Barra}
\affiliation{Dipartimento di Matematica e Fisica ``Ennio De Giorgi'', Universit\`a del Salento, Lecce, Italy
}%
\affiliation{Istituto Nazionale di Fisica Nucleare, Sezione di Lecce, Italy}
\author{Martino Centonze}
\affiliation{Dipartimento di Matematica e Fisica ``Ennio De Giorgi'', Universit\`a del Salento, Lecce, Italy
}%

\author{Alberto Fachechi}
\affiliation{Dipartimento di Matematica e Fisica ``Ennio De Giorgi'', Universit\`a del Salento, Lecce, Italy
}
\affiliation{Istituto Nazionale di Fisica Nucleare, Sezione di Lecce, Italy}
%


\date{\today}

\begin{abstract}
We consider a three-layer Sejnowski machine and show that features learnt via contrastive divergence have a dual representation as patterns in a dense associative memory of order $P=4$. The latter is known to be able to Hebbian-store an amount of patterns scaling as $N^{P-1}$, where $N$ denotes the number of constituting binary neurons interacting $P$-wisely. We also prove that, by keeping the dense associative network far from the saturation regime (namely, allowing for a number of patterns scaling only linearly with $N$, while $P>2$) such a system is able to perform pattern recognition far below the standard signal-to-noise threshold. In particular, a network with $P=4$ is able to retrieve information whose intensity is $\mathcal {O}(1)$ even in the presence of a noise $\mathcal {O}(\sqrt{N})$ in the large $N$ limit. This striking skill stems from a redundancy representation of patterns -- which is afforded given the (relatively) low-load information storage -- and it contributes to explain the impressive abilities in pattern recognition exhibited by new-generation neural networks.
The whole theory is developed rigorously, at the replica symmetric level of approximation, and corroborated by signal-to-noise analysis and Monte Carlo simulations.
\end{abstract}

\pacs{Valid PACS appear here}
\maketitle


Artificial intelligence is nearly everywhere in today's society and has rapidly changed the face of economy, communication and science.
Its global success is mainly due to modern neural-network's architectures which allow deep learning \cite{DL0,HintonLast,Hinton1} and, particularly relevant for the present paper, pattern recognition at prohibitive noise levels \cite{SNR-wall}.
Despite the pervasiveness of such technologies, a clear rationale of the underlying mechanisms is still lacking.
\newline
The Statistical Mechanics of Disordered Systems has been playing a primary role in the theoretical investigation of neural networks since the early studies by  Amit, Gutfreund and Sompolinksy on pairwise associative neural networks \cite{Amit}, and it still constitutes a valuable tool toward an explanation of the impressive skills of modern nets. For instance, recently, Metha $\&$ Schwab have highlighted the profound link between deep learning and renormalization group \cite{Metha}, while Krotov $\&$ Hopfield have showed a duality between higher-order generalizations of the Hopfield model \cite{Hopfield}, referred to as {\em dense associative memories} (DAMs), and neural networks commonly used in (deep) learning \cite{HopfieldKrotov_DAM}, also highlighting remarkable properties of these networks (e.g., the minima of the cost function are devoid of rubbish representations and adversarial patterns fail to fool the dense network) \cite{HopfieldKrotov_DL}.

\begin{figure}
\begin{center}
\includegraphics[scale=0.23]{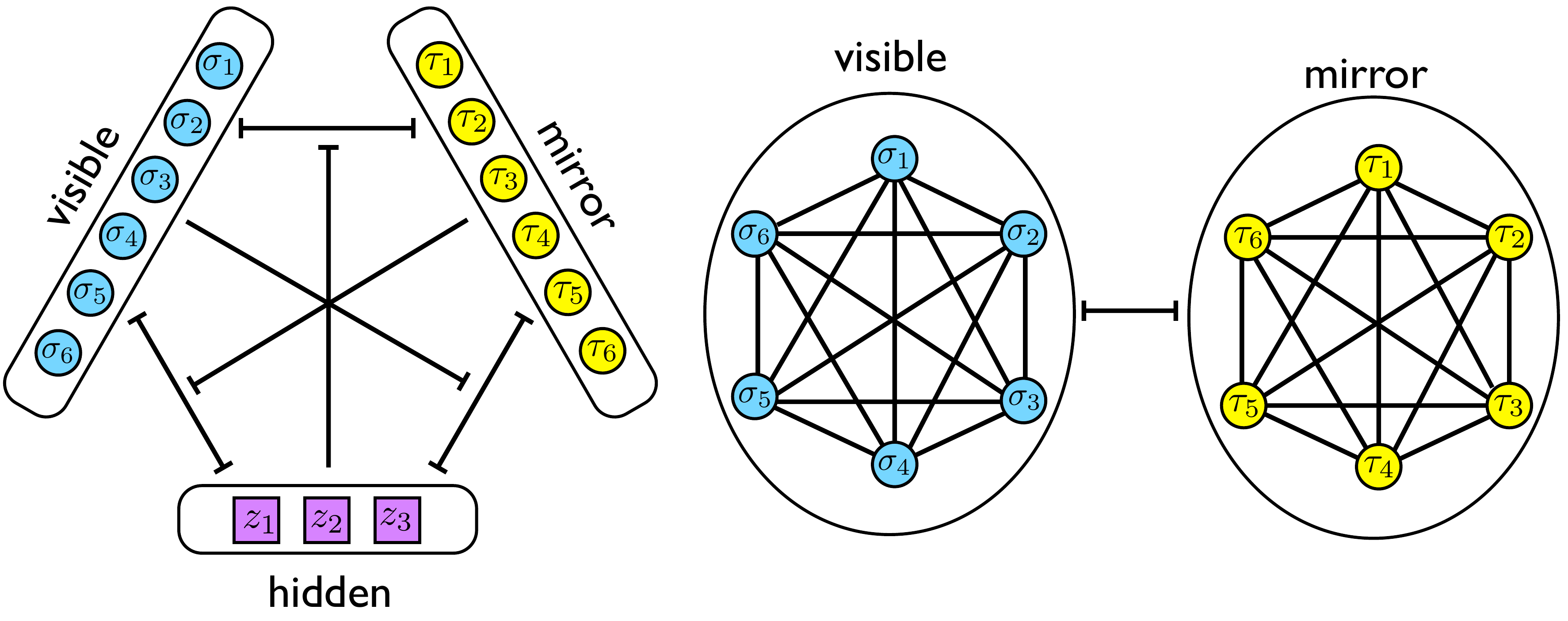}
\caption{Schematic representations of the Restricted Seinowskj Machine (left) and its dual representation in terms of a bipartite Dense Associative Network (right).
In the former, neurons $i, \mu, \rho$ interact 3-wisely through the coupling $\xi_{i \mu}^{\rho}$ (see also eq.~\ref{eq:Hsej}), while, in the latter, neurons $i, \mu, j, \nu$ interact 4-wisely through  the coupling $J_{i \mu}^{j \nu}$ (see also eq.~\ref{eq:Heff}).}
\label{Fig:Uno}
\end{center}
\end{figure}

Here we consider a basic architecture for machine learning, i.e. the \emph{restricted Sejnowski machine} (RSM) \cite{Semio}, that is a third-order Boltzmann machine \cite{Hinton1}, where triples of units interact symmetrically; in the jargon of Statistical Mechanics, this is just a three-layer spin-glass with ($P$=3)-wise interactions. In particular, we equip this network with a standard hidden layer and with two visible layers (a primary and a mirror channel, see Fig.\ref{Fig:Uno} left), which possibly mimic the typical presence of two input sources in biological networks (i.e., the {\em eyes}). As we show, the RSM displays, as a dual representation, a bipartite DAM, i.e., a bipartite Hopfield model with ($P$=4)-wise interactions, see Fig.~\ref{Fig:Uno} right. In this dual representation, the $K$ features embedded in the RSM correspond to the $K$ patterns stored in the DAM. 
\newline
It is worth recalling that a neural network with ($P>2$)-wise interactions among its units does not need to fulfil Gardner's bound: the latter holds solely for quadratic cost functions and implies that, being $K_{\textrm{max}}(N)$ the largest number of random i.i.d. patterns that a network built of $N$ binary neurons can store, then $\lim_{N \to \infty} K_{\textrm{max}}(N)/N = \alpha_c <2$ \cite{Gardner}. In fact, Baldi $\&$ Venkatesh  \cite{Baldi} proved that, for a $P$-spin associative memory built of $N$ binary neurons, $K_{\textrm{max}}(N)  \propto N^{P-1}$ (a result made rigorous by Bovier $\&$ Niederhauser \cite{BovierPspin}); clearly for $P=2$ we recover the standard Hopfield scenario.
\newline
In the last decades, the quest for enhanced storage capacities has strongly  biased the statistical mechanical investigations, possibly limiting alternative inspections of the computational capabilities of these networks, which is the main focus of this work, as summarized hereafter.

In the standard Hopfield model it is possible to retrieve a number $K$ of patterns that is extensive in $N$ (i.e., $K = \alpha N$ with $\alpha \leq 0.14$) by pushing the signal-to-noise ratio to its limit, namely by letting the magnitude $\mathcal {S}$ of the signal -- stemming from the pattern to be retrieved  -- and the magnitude $\mathcal {N}$ of the (quenched) noise -- stemming from the remaining patterns providing an intrinsic glassiness -- share the same order.
Should the information encoded by patterns be affected by some source of noise, the condition $\mathcal {S}/\mathcal {N} \sim \mathcal {O}(1)$ would be deranged in favour of the noise and retrieval capabilities would be lost. On the other hand, as we show, if we let dense ($P=4$) networks operate with a load $K = \alpha N$ (with $\alpha>0$), these turn out to be able to retrieve the information ($\sim \mathcal {O}(1)$) encoded by patterns is perturbed by extensive noise ($\sim \mathcal {O}(\sqrt N)$). This is ultimately due to the possibility of redundant representation of patterns \cite{Redu1,Redu2}, which implies a storage cost of $\mathcal {O}(N^2)$ bits per pattern.
\newline
In the following we give more technical details to prove the previous statements.
%

The RSM \cite{Semio} considered here is built on three layers, two of which -- referred to as {\em visible} and {\em mirror}, respectively (see Fig.\ref{Fig:Uno}, left panel) -- are digital and made up of $N$ Ising neurons per layer, $\boldsymbol{\sigma}\in \{-1,+1\}^{N}$ and $\boldsymbol{\tau} \in \{- 1, +1\}^{N}$, while the third layer -- referred to as {\em hidden} -- is analog and made of $K$ neurons $\boldsymbol{z}$, whose states are i.i.d. Gaussians $\mathcal {N}(0,\beta^{-1})$ ($\beta>0$ tuning the level of the fast noise in the net \cite{Amit}). The model presents third-order interactions among neurons of different layers but no intra-layer interactions (whence the {\em restriction}). Its cost function $H_{\textrm{RSM}}$ is given by
\begin{align}
H_{\textrm{RSM}}(\boldsymbol{\sigma},\boldsymbol{\tau}, \boldsymbol{z} | \boldsymbol{\xi})=-\frac{1}{N^{3/2}}\sum_{i,\mu,\rho=1}^{N,N,K} \xi^{\rho}_{i\mu} \sigma_i \tau_{\mu} z_{\rho},
\label{eq:Hsej}
\end{align}
with $i, \mu=1,..,N$ and $\rho=1,..,K$. In the thermodynamic limit each layer size diverges such that $\lim_{N \to \infty} K/N =\alpha >0$ and the factor $N^{-3/2}$ keeps the mean value of the cost function (under the quenched Gibbs measure \cite{Guerra}) linearly extensive in $N$. The interaction between each triplet of neurons is encoded in the $K \times N \times N$ tensor $\boldsymbol{\xi}$ whose $\rho$-th element will be written as
\begin{align}
\label{eq:factor}
\xi^{\rho}_{i \mu}=\xi^{\rho}_{i} \xi^{\rho}_{\mu}, ~~ i,\mu=1,...,N,
\end{align}
where $\xi^{\rho}_{i} \in \{ -1, +1\}$ is meant as the $i$-th entry of the $\rho$-th pattern to be retrieved in the dual bipartite DAM. Notice that the factorization (\ref{eq:factor}) ensures the symmetry of $\xi^{\rho}_{i \mu}$ for any $\rho$ and it lies at the core of the pattern redundancy scheme pursued here. In fact, the information contained into a set of $K$ binary patterns of length $N$ is inflated into a symmetric tensor of size $K N^2$.
\newline
Given a small learning rate $\epsilon >0$, we obtain for this network the following  contrastive-divergence \cite{ContrDiv} learning rule (see the Appendix A of the SI for details on its derivation and performances)
\begin{align}
\label{eq:contrastive}
\Delta \xi_{i\mu}^\rho= \epsilon \beta \left(\langle  \sigma_i \tau_\mu z_\rho\rangle_+ - \langle  \sigma_i \tau_\mu z_\rho \rangle_-\right),
\end{align}
where the subscript ``$+$'' means that both visible and mirror layers are set at the data input (i.e., they are {\it clamped}), while the subscript ``$-$'' means that all neurons in the network are left free to evolve; importantly, while clamped, visible and mirror layers are always exposed to the same information  (i.e., $\boldsymbol{\sigma} = \boldsymbol{\tau} = \boldsymbol{\xi}^{\rho}$).
\newline
Using the symbol $Dz_{\rho}$ to denote the Gaussian measure with variance $\beta^{-1}$ (i.e.,  $Dz_{\rho} \equiv dz_\rho \exp (-\beta z_{\rho}^2/2) \sqrt{\beta/ 2\pi}$), the partition function $Z$ related to the cost function (\ref{eq:Hsej}) reads
\begin{align}\label{partition}
Z=\sum_{  \boldsymbol{\sigma}, \boldsymbol{\tau}} \int \prod_{\rho=1}^{K} Dz_{\rho} \exp\left(\frac{\beta}{N^{3/2}}\sum_{i,\mu,\rho=1}^{N,N,K} \xi^{\rho}_{i\mu} \sigma_i \tau_{\mu} z_{\rho}\right).
\end{align}
By construction, 
the couplings are symmetric ($\xi_{i \mu}^{\rho} = \xi_{\mu i}^{\rho}$) and detailed balance ensures that the long term relaxation of any (not-pathological) neural dynamics is described by the related Gibbs measure \cite{Amit,BarraEquivalenceRBMeAHN}. Marginalizing over the hidden layer,
\begin{align}
\nonumber
P(\boldsymbol{\sigma},\boldsymbol{\tau} | \boldsymbol{\xi})=\frac{\int D\boldsymbol{z} e^{-\beta H_{\textrm{RSM}}(\boldsymbol{\sigma},\boldsymbol{\tau},\boldsymbol{z} | \boldsymbol{\xi})}}{Z} \equiv \frac{e^{-\beta H_{\textrm{DAM}} (\boldsymbol{\sigma},\boldsymbol{\tau} | \boldsymbol{\xi})}}{Z},
\end{align}
where the last equation tacitly defines the cost function of the DAM, namely
\begin{eqnarray}
\nonumber
H_{\textrm{DAM}}(\boldsymbol{\sigma},\boldsymbol{\tau}| \boldsymbol{\xi})&=&-\frac{1}{2 N^3}\sum_{\rho=1}^K \left(\sum_{i,\mu=1}^{N,N} \xi^{\rho}_{i\mu} \sigma_i \tau_{\mu}\right)^2 \\
&=&-\frac{1}{2 N^3} \sum_{i,j=1}^{N,N} \sum_{\mu,\nu=1}^{N,N} J_{i \mu}^{j \nu} \sigma_i \sigma_j \tau_{\mu} \tau_{\nu},
\label{eq:Heff}
\end{eqnarray}
where $J_{i \mu}^{j \nu}=\left(\sum_{\rho} \xi^{\rho}_{i\mu} \xi^{\rho}_{j\nu} \right)$. This decomposition shows that the $\boldsymbol {\xi}$'s play as eigenvectors for the tensor $\boldsymbol J$, whose symmetry with respect to an exhange of indices $(i,\mu)$ and $(j,\nu)$ mirrors the symmetry between the $\boldsymbol \sigma$ and the $\boldsymbol \tau$ variables underlying the learning rule (\ref{eq:contrastive}). Notice that $H_{\textrm{DAM}}$ corresponds to a ($P$=4)-wise bipartite Hopfield model (see Fig.~\ref{Fig:Uno}, right panel), namely a minimal generalization of the Hebbian kernel in the classic Hopfield reference (quite similar to auto-encoders in Engineering jargon \cite{HintonAutoEncoder}). Also, this equivalence generalizes the standard duality between restricted Boltzmann machines and (pairwise) Hopfield neural networks \cite{Agliari-Dantoni,BarraEquivalenceRBMeAHN}.
\newline
To start dealing with network's capabilities, it is convenient to introduce generalized Mattis order parameters $M_{\rho}$ defined as
\begin{align}
\label{eq:Mrho}
M_{\rho} \equiv \frac{1}{N^2}\sum_{i,\mu=1}^{N,N} \xi^{\rho}_{i\mu} \sigma_i \tau_{\mu}.
\end{align}
The signal-to-noise analysis for this system can be obtained by requiring the dynamic stability of the neural state recalling, without loss of generality, the pattern $\rho=1$, that is, $\sigma_i \tau_{\mu} = \xi_{i \mu}^1$. Therefore, denoting with $h_{i \mu}$ the internal field acting on $\sigma_i$ and $\tau_{\mu}$ we get
\begin{eqnarray}
\nonumber
\sigma_i \tau_{\mu} h_{i \mu} &=& \mathcal {S} + \mathcal {N} = \frac{1}{2N} \sum_{\rho=1}^K M_{\rho} \xi_{i \mu}^{\rho} \xi_{i \mu}^{1} \\
\label{eq:S2N}
&=&   \frac{1}{2N}  \left[ M_1 + \sum_{\rho>1}^K M_{\rho} \xi_{i \mu}^{\rho} \xi_{i \mu}^{1}  \right].
\end{eqnarray}
As the signal term inside the brackets in (\ref{eq:S2N}) is $M_1 \sim \mathcal {O}(1)$, while the noise term corresponds to a sum of $(K-1)$ stochastic and uncorrelated contributions, each of order $\mathcal {O}(N^{-1})$, exploiting the central limit theorem it is immediate to check that the quenched noise due to non-retrieved patterns can be amplified by a factor $\sqrt{N}$ still preserving the stability condition $\mathcal {S}/\mathcal {N} \sim \mathcal {O}(1)$. We can therefore introduce noisy patterns yielding to the noisy tensor $\boldsymbol{\eta}$ with entries
\begin{align}
\label{eq:noisypat}
\eta^\rho_{i\mu} \equiv  \xi_{i\mu}^\rho +\sqrt K \tilde{\xi}_{i\mu}^\rho,
\end{align}
where the information is carried by the Boolean entries of ${\xi}_{i\mu}^\rho$, while the noise is coded in the real $\tilde{\xi}_{i\mu}^\rho$ that are  i.i.d. standard Gaussian variables for $i,\mu=1,...,N$ and $\rho=1,...,K$. Notice that the information encoded by the patterns is perturbed by adding a stochastic term $\tilde{\boldsymbol{\xi}}$ on $\boldsymbol{\xi}$ (eq.\ref{eq:noisypat}) rather than directly on $\boldsymbol J$; the latter choice would have a lower impact on network capacity and is therefore less challenging.
%
%
%
In analogy with (\ref{eq:Mrho}) we also define
\begin{align}\label{eq:magnetization}
\tilde M_\rho \equiv \frac1{N^2}\sum_{i,\mu=1}^{N,N} \tilde{\xi}^\rho _{i\mu}\sigma_i \tau_\mu. 
\end{align}
Replacing the Boolean tensor (\ref{eq:factor}) in eq.~(\ref{eq:Heff})  with the noisy tensor (\ref{eq:noisypat}) and exploiting the definitions (\ref{eq:Mrho}) and (\ref{eq:magnetizations}), we get $H_{\textrm{DAM}}=- \frac{N}{2} \sum_{\rho} (M_{\rho} +\sqrt{K} \tilde{M}_\rho)^2$.
Then, in the limit of large $N$, splitting the signal and the noise contributions, the Boltzmann factor in eq. (\ref{partition}) reads as (see also Appendix B in the SI)
\begin{align}
\nonumber
\exp\left(-\beta H_{\textrm{RSM}}\right) \underset{N\to \infty}{\sim} \exp\left(\beta \frac{N}{2} M^2_1 + \beta \frac{\alpha N^2}{2}  \sum_{\rho\geq 2}^K \tilde{M}^2_\rho\right).
\end{align}
%
\begin{figure} [bt]
\begin{center}
\includegraphics[scale=0.3]{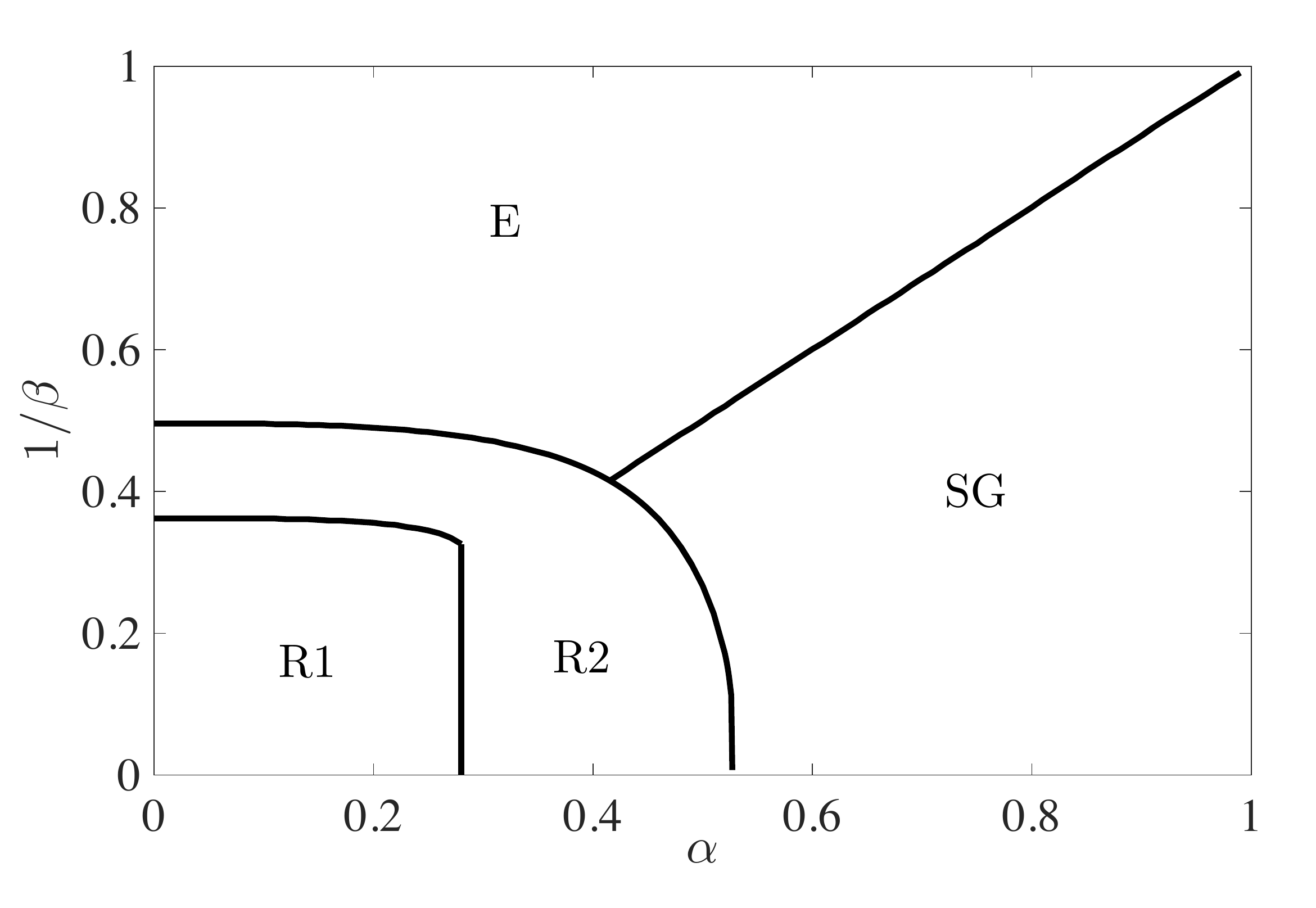}
\caption{Phase diagram for the DAM with $(P=4)$-wise interactions among the $N$ neurons and a load $K= \alpha N$, as a function of the capacity $\alpha$ and of the the noise level $1/\beta$. This diagram was obtained by solving the self-consistent equations (\ref{eq:SC0})-(\ref{eq:SC2}) and by identifying the retrieval region as the region where each neural configurations corresponding to the stored patterns (and their symmetric version) is a maximum of the pressure -- either global, (R1) or local (R2) -- the spin-glass (SG) region as the region where retrieval capabilities are lost due to prevailing ``slow noise'' $\alpha$, and the ergodic (E) region as the region where retrieval capabilities are lost due to prevailing ``fast noise'' $1/\beta$ (see Appendix D for further details).}
\label{Fig:Due}
\end{center}
\end{figure}	
%
Let us now handle the two terms appearing as argument of the exponential in the r.h.s.: exploiting the redundancy $\xi^{1}_{i\mu}=\xi_i^1 \xi_\mu^1$ and calling $m_{\sigma}$ and  $m_{\tau}$ the Mattis magnetization related to the visible layer $\boldsymbol\sigma$ and to the mirror layer $\boldsymbol\tau$ respectively, we get $ M_{1}=\left(\frac{1}{N}\sum_{i} \xi_i^1 \sigma_i\right) \left(\frac{1}{N}\sum_{\mu}\xi_\mu^1 \tau_{\mu}\right)\equiv m_{\sigma} m_{\tau}$, in such a way that $\beta N M^2_1 /2 = \beta N m^2_{\sigma}m^2_{\tau}/2$; by performing a Hubbard-Stratonovich transformation, the quenched noise given by the non-retrieved $K-1$ patterns is linearized as $\sqrt{\alpha \beta} \sum_{i,\mu,\rho\geq 2} \tilde{\xi}^{\rho}_{i\mu} \sigma_i \tau_\mu z_\rho/N$.
After these passages one can address the evaluation of the intensive quenched pressure of the model, defined as,
\begin{align}
\nonumber
A(\alpha, \beta) \equiv \lim_{N \to \infty} \frac{1}{N} \mathbb{E}_{\boldsymbol{\eta}} \ln \sum_{\boldsymbol{\sigma},\boldsymbol{\tau}} \int \prod_{\rho=1}^{K} Dz_\rho \exp\left(-\beta H_{\textrm{RSM}}\right),
\end{align}
exploiting Guerra's interpolation techniques \cite{BarraEquivalenceRBMeAHN,Agliari-Barattolo}. Under the Replica Symmetric (RS) ansatz, the quenched pressure reads as (see Appendix C in the SI for technical details)
\begin{align} \nonumber
A^{\textrm{RS}}=& 2\ln 2+\frac{\alpha^2 \beta^2}{2}p(2  q r - r - q) - \frac{3}{2}\beta \overline{m}^2_{\sigma} \overline{m}^2_{\tau}  \\ \nonumber
&+ \int Dx \ln \cosh\left( \alpha \beta x\sqrt{r p}   + \beta  \overline{m}_{\sigma} \overline{m}_{\tau}^2 \right) \\ \nonumber
&+ \int Dx \ln \cosh\left(\alpha \beta  x\sqrt{  q p}  + \beta  \overline{m}^2_{\sigma} \overline{m}_{\tau} \right) \\
\label{eq:ARS}
&-\frac{\alpha}{2} \ln [1-\alpha\beta  (1-q r)]+\frac{\alpha^2 \beta}{2} \frac{q r}{1-\alpha\beta  (1-q r)},
\end{align}
where $\overline m_\sigma$ and $\overline m_\tau$ are the RS values of  the Mattis magnetizations, while $q$, $p$ and $r$ are the RS values for the two-replica overlaps for each layer (visible, hidden and mirror respectively).
Its extremization returns the following self-consistency equations for the order parameters
\begin{eqnarray}
\label{eq:SC0}
&&q=\int Dx \tanh^2\left(\alpha\beta \sqrt{ rp} x + \beta  \overline{m}_{\sigma}\overline{m}^2_{\tau} \right),\\
&&r=\int Dx \tanh^2\left(\alpha\beta \sqrt{  qp} x + \beta \overline{m}^2_{\sigma} \overline{m}_{\tau} \right),\\
&&p= \frac{\alpha q r}{\left[1-\alpha\beta (1-qr)\right]^2},\\
\label{eq:SC1}
&&\overline{m}_{\sigma}=\int Dx \tanh \left(\alpha\beta \sqrt{  rp} x + \beta  \overline{m}_{\sigma} \overline{m}^2_{\tau} \right),\\
\label{eq:SC2}
&&\overline{m}_{\tau} =\int Dx \tanh \left(\alpha\beta \sqrt{ qp} x + \beta  \overline{m}^2_{\sigma}\overline{m}_{\tau}\right).
\end{eqnarray}
whose solution paints the phase diagram in Fig. \ref{Fig:Due} (see Appendix D in the SI for more details).
%
%
%
%
\begin{figure}
\begin{center}
\includegraphics[scale=0.3]{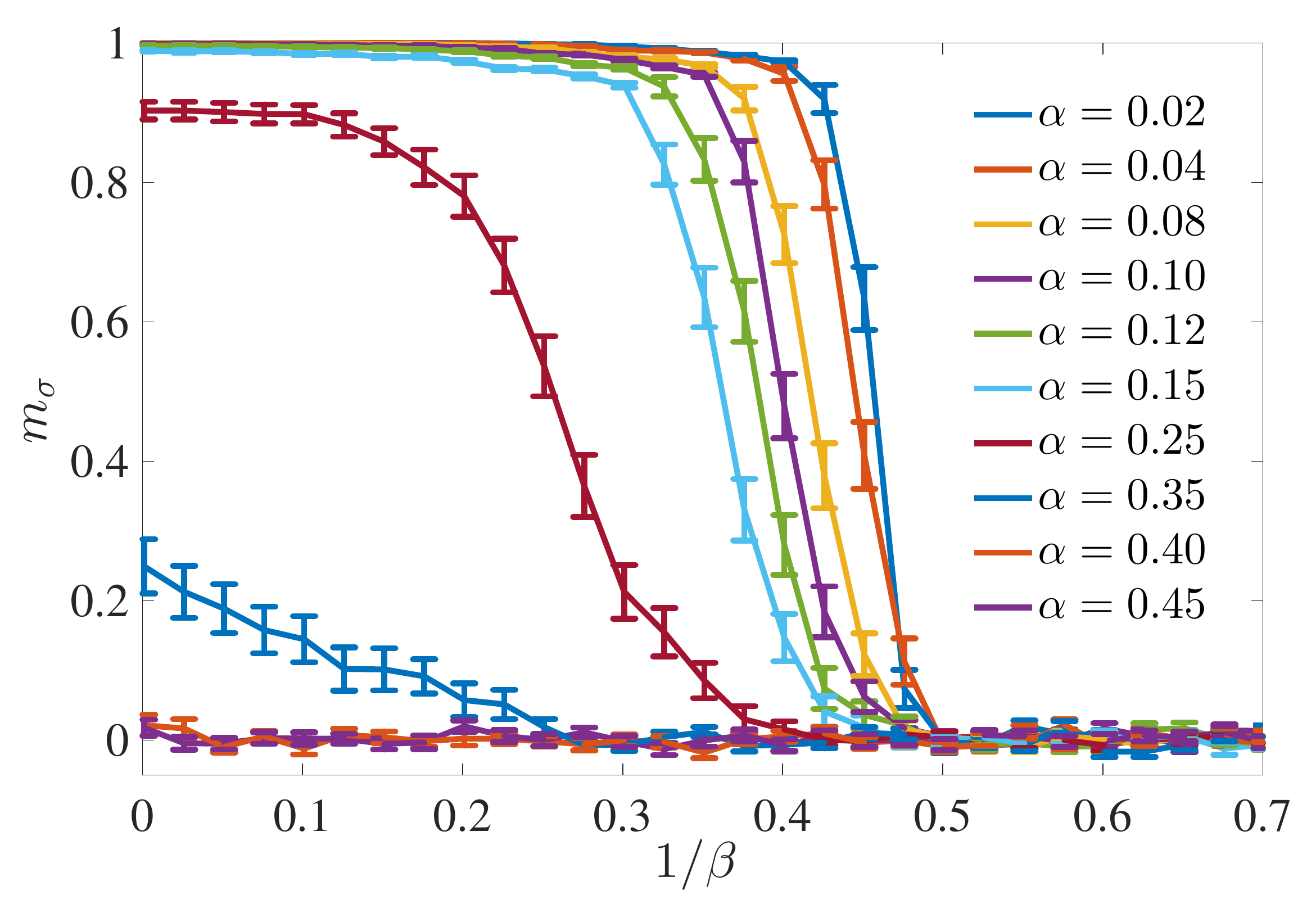}
\caption{Expected Mattis magnetization obtained from Monte Carlo simulations run for $N=150$ and for different values of $\alpha$, as a function of $1/\beta$}. Notice that, as $\alpha$ is tuned from $0.25$ to $0.35$, the magnetization abruptly drops even at small values of $1/\beta$, consistently with the transition from the region R1 to the region R2 found theoretically (see Fig.~\ref{Fig:Due} and Appendix E in the SI for further details and discussions).
\label{Fig:Tre}
\end{center}
\end{figure}	
\newline
The theory is also corroborated via Monte Carlo simulations; a sample of this analysis is shown in Fig.~\ref{Fig:Tre}, while more extensive discussions can be found in Appendix E of the SI.
%

To summarise, we considered a Sejnowski machine equipped with two visible layers and we showed that it can perform pattern-redundant representation via a suitable generalization of the standard contrastive divergence. Further, we proved that this machine has a dual representation in terms of a bipartite DAM in such a way that the features learnt by the former correspond to the patterns stored in the latter and, whatever the learning mode (adaptive versus Hebbian), in the operational mode these networks achieve pattern recognition always in a Hebbian fashion. We studied these nets via statistical mechanical tools obtaining (under the RS ansatz) a phase diagram, where their remarkable capabilities shine. In particular, there exists a region in the parameter space where they can retrieve patterns although these are (apparently) overpowered by the noise. This may contribute to explain the high-rate ability of deep/dense networks in pattern recognition, as empirically evidenced in a variety of tasks. Indeed, at finite volumes (as standard dealing with real data-sets), it is not obvious which regime of operation the network is actually set at: to see this one can notice that at finite $N$ and  $K$ one has only access to the ratio $\alpha(K,N) = K/N$ which can possibly be compatible with different scalings (e.g., $K = \alpha_1 N^{P-1}$ or $K = \alpha_2 N$).
Hence, we speculate that such impressive detection skills emerge when these nets are away from the memory storage saturation.  Further, we have shown by a pure statistical mechanical perspective, how pattern recognition power and memory storage are strongly related.
\newline
For the sake of completeness, we report that also in the purely engineering counterpart, pattern redundancy is exploited to cope with high noise rate (e.g., in white Gaussian additive channels  \cite{WGAC,DirSeq}).  In particular, our approach is close to the so called {\em channel access method} in telecommunications, namely a set-up where more than two terminals connected to the same transmission medium  are allowed to share its capacity.
%

\appendix
\section{Derivation and performances of the Contrastive-Divergence learning rule}

\begin{figure}
	\begin{center}
		\includegraphics[scale=0.4]{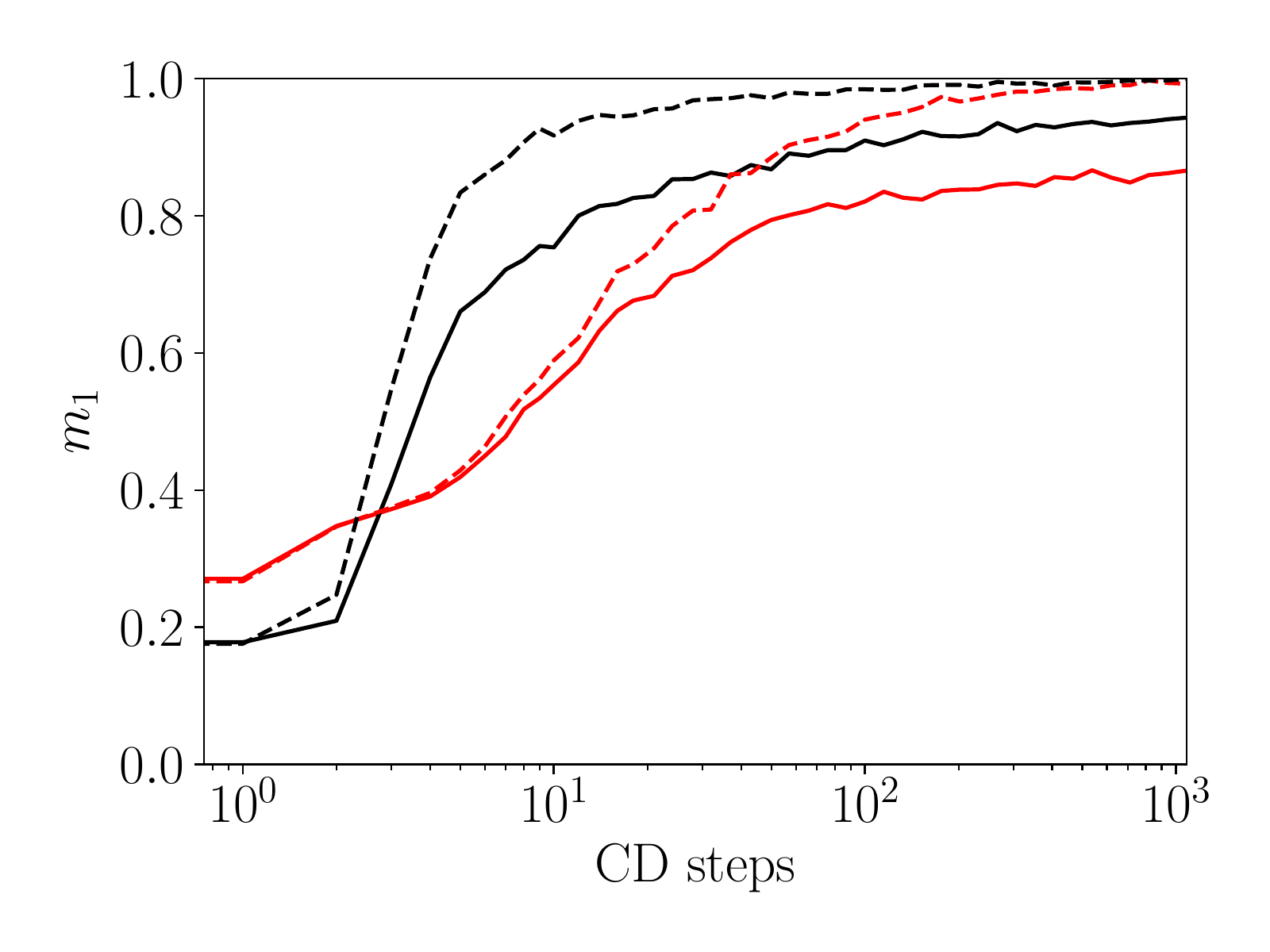}
		\includegraphics[scale=0.4]{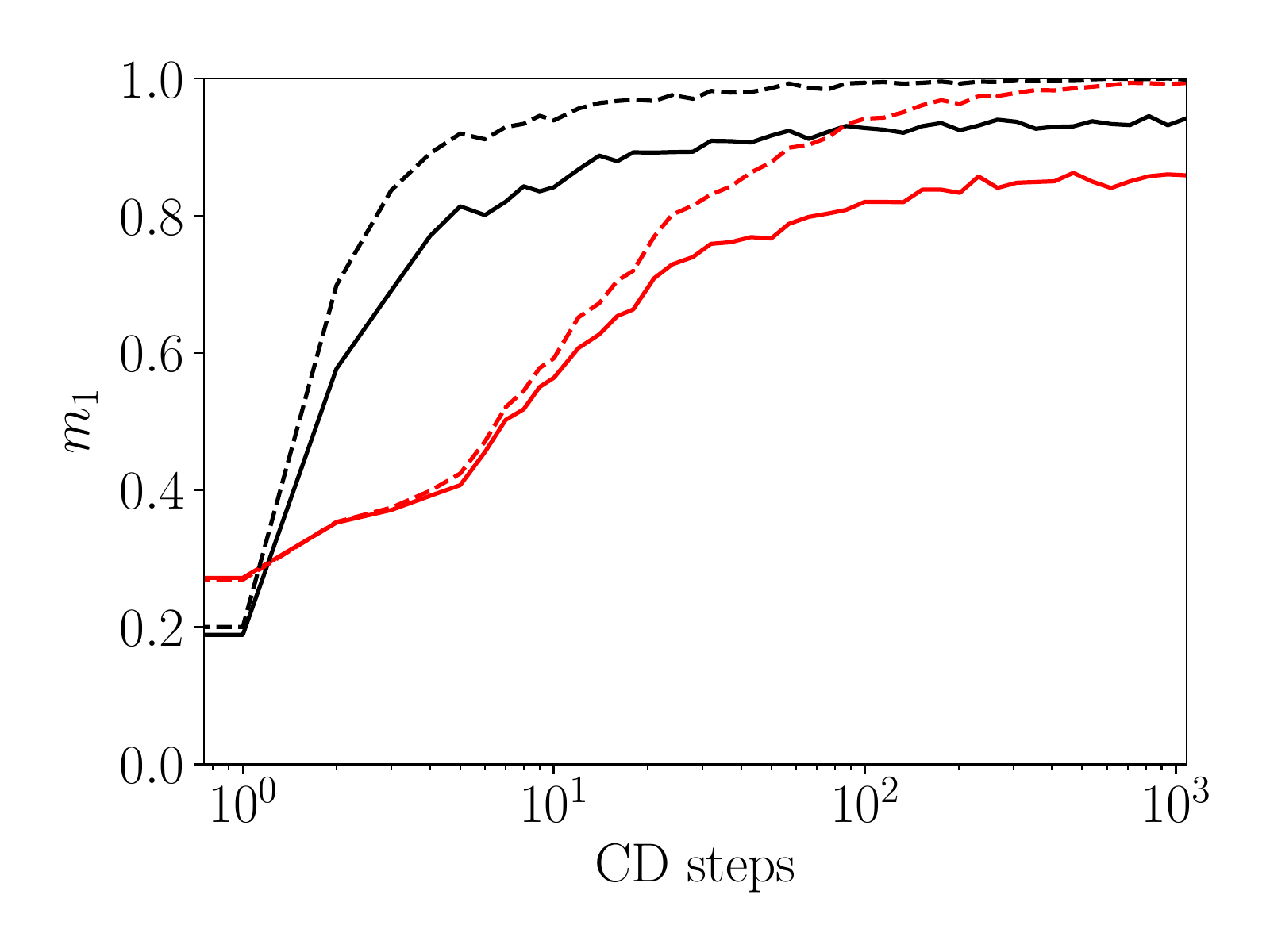}
		\caption{The two plots show a comparison between learning performances of a RSM (black lines) and RBM (red lines). 
				Dashed lines are for comparison of the performances of the machines during the training stage while solid lines are for comparison during the validation stage. On the horizontal axes, we report the number of CD-steps while on the vertical axes we show the overlap between the visible layer $\boldsymbol{\sigma}$ and the retrieved test-pattern $\boldsymbol{\xi}^1$ (i.e., the magnetization $m_1$). In the left plot the two networks work with the optimal learning rate for the RBM (evaluated as $\epsilon=0.266$), nonetheless the RSM outperforms the RBM both in the training and in the validation stages. In the right plot, the two networks operate with their respective optimal learning rates (that is $\epsilon_{\text{RBM}}=0.266$ and $\epsilon_{\text{RSM}}=0.52$) and the difference in the performances is further enhanced. In both cases, the network size is fixed to $N=20$ (however, we obtained analogous results up to $N=200$). Similar results also hold for the overlap with pattern $\boldsymbol\xi^2$.}
		\label{Fig:performances}
	\end{center}
\end{figure}

Let us recall the partition function (4) of the Sejnowski machine we are inspecting
\begin{align}\nonumber
Z= &\sum_{\boldsymbol \sigma, \boldsymbol \tau} \int \Big(\prod_{\rho=1}^K \frac{dz_\rho}{\sqrt{2\pi \beta^{-1}}}\Big)\times\\
&\times\exp \Big(-\frac{\beta}{2}\sum_{\rho=1}^K z_\rho^2+\beta\sum_{i,\mu,\rho=1}^{N,N,K} \hat \xi_{i\mu}^\rho \sigma_i \tau_\mu z_\rho\Big),
\end{align}
where $ \hat \xi_{i\mu}^\rho =N^{-3/2}\xi_{i\mu}^\rho$. Such expression suggests that learning should act on the couplings $\xi_{i\mu}^\rho$ rather than on the information patterns $\xi^\rho_i$ (as it happens in the simpler pairwise scenario \cite{ContrDiv,BarraEquivalenceRBMeAHN}). In order for the learning procedure to create free-energy (we recall that the pressure $A$ is simply related to the free energy $F$ by $A = - \beta F$, in such a way that the two functions exhibit the same extreme points, yet maxima in the pressure just corresponds to minima in the free-energy) minima placed at $\boldsymbol\sigma=\boldsymbol\tau= \boldsymbol \xi^\rho$, both the visible and mirror layers should be set according to the data, namely the two {\em eyes} of the machine do look at the same outside world.
\newline
In this Appendix, we prove that the learning rule reads as
\begin{align} \label{eq:GD}
\Delta \hat \xi_{i\mu}^\rho= \epsilon \beta \left(\langle  \sigma_i \tau_\mu z_\rho\rangle_+ - \langle  \sigma_i \tau_\mu z_\rho \rangle_-\right),
\end{align}
where the subscript ``$+$'' means that both visible and mirror layers are set at the data input (i.e., they are {\it clamped}), while the subscript ``$-$'' means that all neurons in the network are left free to evolve. Let us write explicitly the probability distribution for a given configuration state:
\begin{align}\nonumber
P(\boldsymbol \sigma,\boldsymbol z ,\boldsymbol \tau)&=Z^{-1} \left(\frac{1}{\sqrt{2\pi \beta^{-1}}}\right)^K \times\\
\times&\exp \Big(-\frac{\beta}{2}\sum_{\rho=1}^K z_\rho^2+\beta\sum_{i,\mu,\rho=1}^{N,N,K} \hat \xi_{i\mu}^\rho \sigma_i \tau_\mu z_\rho\Big),
\end{align}
and suppose the set of data is made of i.i.d. entries generated by a probability distribution $Q(\boldsymbol{\sigma})$, whose features we aim to extract.
\newline
Since the mirror layer, by definition, should mimic the activity of the visible layer (as we want to put the information content in pure minima of the free energy given by configurations of the form $\boldsymbol \tau=\boldsymbol \sigma$), we have to build a representation of the couplings $\hat \xi^\rho_{i\mu}$ such that the marginal distribution $P(\boldsymbol \sigma, \boldsymbol \tau=\boldsymbol \sigma)$ is the best approximation for $Q(\boldsymbol \sigma)$, where
\begin{align}\nonumber
P(\boldsymbol \sigma, \boldsymbol \tau)&=Z^{-1}\int \Big(\prod_{\rho=1}^K \frac{dz_\rho}{\sqrt{2\pi \beta^{-1}}}\Big)\times\\\nonumber
\times&\exp \Big(-\frac{\beta}{2}\sum_{\rho=1}^K z_\rho^2+\beta\sum_{i,\mu,\rho=1}^{N,N,K} \hat \xi_{i\mu}^\rho \sigma_i \tau_\mu z_\rho\Big)\\
&:=\frac{Z(\boldsymbol \sigma,\boldsymbol \tau)}{Z}.
\end{align}
Therefore, we introduce the Kullback-Leibler cross-entropy as
\begin{align}\nonumber
D(P,Q)= \sum_{\boldsymbol\sigma} Q(\boldsymbol\sigma)\log \frac{Q(\boldsymbol\sigma)}{P(\boldsymbol \sigma, \boldsymbol \tau=\boldsymbol \sigma)}.
\end{align}
Under a gradient-descent approach, we have to compute the derivative of the cross-entropy w.r.t. the couplings, that reads as
\begin{align}\nonumber
\label{eq:A6}
\frac{\partial D}{\partial \hat \xi^\rho_{i\mu}}&= -\sum_{\boldsymbol{\sigma}}Q(\boldsymbol \sigma) \times\\&\times\left[ Z(\boldsymbol \sigma, \boldsymbol \tau=\boldsymbol \sigma)^{-1}\frac{\partial Z(\boldsymbol \sigma, \boldsymbol \tau=\boldsymbol \sigma)}{\partial \hat \xi^\rho_{i\mu}}-Z^{-1}\frac{\partial Z}{\partial \hat \xi^\rho_{i\mu}}\right].
\end{align}
The first term in the square brackets of eq. (\ref{eq:A6}) can be written as:
\begin{align}
\begin{split}
Z(\boldsymbol \sigma, \boldsymbol \tau&=\boldsymbol \sigma)^{-1}\frac{\partial Z(\boldsymbol \sigma, \boldsymbol \tau=\boldsymbol \sigma)}{\partial \hat \xi^\rho_{i\mu}}=\\=\,&Z(\boldsymbol \sigma, \boldsymbol \tau=\boldsymbol \sigma)^{-1}\int \Big(\prod_{\rho=1}^K \frac{dz_\rho}{\sqrt{2\pi \beta^{-1}}}\Big)\beta \sigma_i \tau_\mu z_\rho\times\\&\times\exp \Big(-\frac{\beta}{2} \sum_{\rho=1}^K z_\rho^2+\beta \sum_{i, \mu, \rho=1}^{N,N,K} \hat \xi_{i\mu}^\rho \sigma_i \tau_\mu z_\rho\Big)=\\
=\,&Z(\boldsymbol \sigma, \boldsymbol \tau=\boldsymbol \sigma)^{-1}\int \Big(\prod_{\rho=1}^K {dz_\rho}\Big)\beta \sigma_i \tau_\mu z_\rho ~ P(\boldsymbol{\sigma},\boldsymbol{z},\boldsymbol{\tau}=\boldsymbol \sigma),
\end{split}
\label{eq:A7}
\end{align}
and, using Bayes' theorem under the constraint $\boldsymbol \tau=\boldsymbol \sigma$,
\begin{align}\nonumber
\label{eq:A8}
P(\boldsymbol{\sigma},\boldsymbol{z},\boldsymbol{\tau}=\boldsymbol \sigma)&=P(\boldsymbol{\sigma},\boldsymbol{\tau}=\boldsymbol \sigma)P (\boldsymbol z\vert \boldsymbol \sigma, \boldsymbol \tau=\boldsymbol \sigma)=\\&=\frac{Z(\boldsymbol{\sigma},\boldsymbol{\tau}=\boldsymbol \sigma)}{Z}P (\boldsymbol z\vert \boldsymbol \sigma, \boldsymbol \tau=\boldsymbol \sigma).
\end{align}
Therefore, combining (\ref{eq:A7}) and (\ref{eq:A8})
\begin{align}\nonumber
&Z(\boldsymbol \sigma, \boldsymbol \tau=\boldsymbol \sigma)^{-1}\frac{\partial Z(\boldsymbol \sigma, \boldsymbol \tau=\boldsymbol \sigma)}{\partial \hat \xi^\rho_{i\mu}}=\\&=\int \Big(\prod_{\rho=1}^K {dz_\rho}\Big)\beta \sigma_i \tau_\mu z_\rho P (\boldsymbol z\vert \boldsymbol \sigma, \boldsymbol \tau=\boldsymbol \sigma).
\end{align}
When taking the $Q$-weighted sum, we have
\begin{align}\nonumber
&\sum_{\boldsymbol \sigma}Q(\boldsymbol \sigma)Z(\boldsymbol \sigma, \boldsymbol \tau=\boldsymbol \sigma)^{-1}\frac{\partial Z(\boldsymbol \sigma, \boldsymbol \tau=\boldsymbol \sigma)}{\partial \hat \xi^\rho_{i\mu}}=\\\nonumber
&=\sum_{\boldsymbol \sigma}\int \Big(\prod_{\rho=1}^K {dz_\rho}\Big)\beta \sigma_i \tau_\mu z_\rho Q(\boldsymbol \sigma) P (\boldsymbol z\vert \boldsymbol \sigma, \boldsymbol \tau=\boldsymbol \sigma)=\\
&=\beta \langle \sigma_i \tau_\mu z_\rho\rangle_+,
\end{align}
since data are extracted with probability $Q(\boldsymbol \sigma)$.
\newline
The second term in the square brackets of eq.~(\ref{eq:A6}) can be written as:
\begin{align}
Z^{-1}\frac{\partial Z}{\partial \hat \xi^\rho_{i\mu}}= \sum_{\boldsymbol \sigma',\boldsymbol \tau}\int \Big(\prod_{\rho=1}^K {dz_\rho}\Big)\beta \sigma'_i \tau_\mu z_\rho  ~ P(\boldsymbol{\sigma'},\boldsymbol{z},\boldsymbol{\tau}),
\end{align}
where now there are no constraints on the mirror and visible layers. Since there is no dependence on $\boldsymbol \sigma$, the $Q$-weighted sum can be trivially performed, as\\$\sum_{\boldsymbol \sigma}Q(\boldsymbol \sigma)=1$, leading to
\begin{align}\nonumber
&\sum_{\boldsymbol \sigma}Q(\boldsymbol \sigma)Z^{-1}\frac{\partial Z}{\partial \hat \xi^\rho_{i\mu}}=\\\nonumber
&= \sum_{\boldsymbol \sigma',\boldsymbol \tau}\int \Big(\prod_{\rho=1}^K {dz_\rho}\Big)\beta \sigma'_i \tau_\mu z_\rho ~ P(\boldsymbol{\sigma'},\boldsymbol{z},\boldsymbol{\tau})=\\
&=\beta \langle \sigma_i \tau_\mu z_\rho\rangle_-.
\end{align}
All together, we have
\begin{align}
\frac{\partial D}{\partial \hat \xi^\rho_{i\mu}}=-\beta( \langle \sigma_i \tau_\mu z_\rho\rangle_+- \langle \sigma_i \tau_\mu z_\rho\rangle_-).
\end{align}
The gradient descent rule (\ref{eq:GD}) can therefore be expressed in a contrastive divergence (CD) form as $\Delta \hat \xi^{\rho}_{i\mu}=-\epsilon \frac{\partial D}{\partial \tilde \xi^\rho_{i\mu}}$.
In the formula (3) presented in the main text, the factor $N^{3/2}$ was absorbed into the constant $\epsilon$.
\newline
In order to check the performance of this network we proceed as follows: we consider the Restricted Sejnowski Machine (RSM) and, for comparison, a standard Restricted Boltzmann Machine (RBM) and, for both the networks, we arbitrarily choose two random configurations ($\boldsymbol \xi^1$, $\boldsymbol \xi^2$) to be the patterns to be learnt. Via Gibbs-sampling we generate a training set (the same for both the networks) by producing corrupted versions of these patterns (with a level of corruption up to $30\%$). The latter are thus learnt simultaneously via CD and, once the training stage is over, pattern retrieval is further examined. The overlaps $m_{1,2}$ are measured and compared in the training and in the validation stages. In all the tests we performed -- a sample of which is shown in Fig.~\ref{Fig:performances} -- the RSM outperforms the standard RBM (all the tests produced results similar to those reported in Fig.~\ref{Fig:performances}). In particular, beyond being more accurate, the CD-algorithm for the RSM is significantly faster with respect to its RBM counterpart, that is, it reaches large values of $m_{1,2}$ already for a relatively small number of CD steps.\\
	As a last remark we notice that, in the very initial stage (when the number of CD steps is small), the RBM displays a large overlap with respect to the RSM. This effect is of purely stochastic nature as the RBM is fed with a vector of $N$ entries while the RSM is fed with a matrix of $N^2$ entries, in such a way that a random initial configuration will exhibit a larger alignment in the former case. This remark further highlights the higher speed of the RSM.

\section{Signal to noise stability analysis}
In this Appendix, we perform a signal to noise analysis \cite{Amit} for the Dense Associative Memory (DAM) with $(P=4)$-wise interactions among spins and in the linear storage regime $K = \alpha N, \alpha >0$. Its Hamiltonian, or {\em cost function} to keep a Machine Learning jargon, appearing in eq.~(5) in the main text, can be rewritten as
\begin{eqnarray} \label{eq:DAM_SI}
H_{\textrm{DAM}}=-\sum_{i,\mu=1}^{N,N}h_{i\mu}\sigma_i \tau_\mu,
\end{eqnarray}
where
\begin{align}
\label{eq:intf}
h_{i\mu}=\frac1{2N^3}\sum_{j,\nu=1}^{N,N}\sum_{\rho=1}^{K}\eta^\rho _{i\mu}\eta^\rho_{j\nu} \sigma_j  \tau _\nu,
\end{align}
are the internal fields acting on the dimer $\sigma_i \tau_\mu$.\\
The tensors $\boldsymbol{\eta}$ (see also eq.\ref{eq:noisypat}) in the main text and \cite{Barbier}) are expressed as the sum of a Boolean contribution $\boldsymbol{\xi}$ providing the signal and a real contribution $\tilde{\boldsymbol{\xi}}$ accounting for a noise source:
\begin{align}
\eta^\rho_{i\mu}= \xi_{i\mu}^\rho +\sqrt K \tilde{\xi}_{i\mu}^\rho=\xi_{i\mu}^\rho+\sqrt{\alpha N}\tilde{\xi}_{i\mu}^\rho,
\end{align}
where $\mathbb P(\xi_{i\mu}^\rho=\pm1)=1/2$ and $\mathbb P(\tilde{\xi}_{i\mu}^\rho)=\mathcal  N(0,1)$ for each $i,\mu=1,\dots, N$ and $\rho=1,\dots,K$.
\newline
We recall the definition of the $2K$ generalized Mattis magnetizations as
\begin{eqnarray}\label{eq:magnetizations}
M_\rho&=\frac1{N^2}\sum_{i,\mu=1}^{N,N} \xi^\rho _{i,\mu}\sigma_i \tau_\mu, \\
\tilde M_\rho&=\frac1{N^2}\sum_{i,\mu=1}^{N,N} \tilde{\xi}^\rho _{i\mu}\sigma_i \tau_\mu, 
\end{eqnarray}
with $\rho=1,...,K$.
In terms of these overlaps, the internal fields (\ref{eq:intf}) can be written as
\begin{align}\nonumber
h_{i\mu}=& \frac{1}{2N^3}\sum_{j,\nu=1}^{N,N}\sum_{\rho=1}^K \left(\xi_{i\mu}^\rho +\sqrt{\alpha N} \tilde{\xi}_{i\mu}^\rho \right) \left(\xi_{j\nu}^\rho +\sqrt{\alpha N} \tilde{\xi}_{j\nu}^\rho \right)\sigma_j \tau_\nu
\\\nonumber
=&\frac{1}{2N^3}\sum_{j,\nu=1}^{N,N}\sum_{\rho=1}^K \Big( \xi_{i\mu}^\rho\xi_{j\nu}^\rho \sigma_j\tau_\nu+\sqrt{\alpha N}\xi_{i\mu}^\rho \tilde{\xi}^\rho_{j\nu}\sigma_j\tau_\nu+\\\nonumber
&\qquad+\sqrt{\alpha N} \tilde{\xi}^\rho_{i\mu} \xi^\rho_{j\nu}\sigma_j\tau_\nu+\alpha N \tilde{\xi}^\rho_{i\mu} \tilde{\xi}^\rho_{j\nu}\sigma_j\tau_\nu \Big)\\\nonumber
=&\frac1{2N}\sum_{\rho=1}^K \Big(\xi^\rho_{i\mu}M_\rho +\sqrt{\alpha N} \xi^\rho_{i\mu}\tilde M_\rho+\\
&\qquad+\sqrt{\alpha N} \tilde{\xi}^\rho_{i\mu}M_\rho+\alpha N \tilde{\xi}^\rho_{i\mu}\tilde M_\rho \Big).
\end{align}
We aim to check the stability of configurations where the dimers $\sigma_i \tau_{\mu}$ are aligned to a given element of the tensor, say $\xi^1 _{i\mu}$; the resulting contribution to the energy function \eqref{eq:DAM_SI} is
\begin{align}
\nonumber
h_{i\mu} \xi^1_{i\mu}&=\frac1{2N}\sum_{\rho=1}^K(\xi^\rho_{i\mu}\xi^1 _{i\mu}M_\rho +\sqrt{\alpha N}\xi^\rho_{i\mu}\xi^1 _{i\mu}\tilde M_\rho+\\\nonumber
&\qquad+\sqrt{\alpha N} \tilde{\xi}^\rho_{i\mu}\xi^1 _{i\mu}M_\rho+\alpha N \tilde{\xi}^\rho_{i\mu}\xi^1 _{i\mu}\tilde M_\rho)\\
\label{eq:stability}
&= \frac1{2N}\Big[M_1+\sqrt{\alpha N} \tilde{\xi}^1_{i\mu}\xi^1 _{i\mu}M_1+ \\\nonumber
&\qquad+\sum_{\rho=2}^K(\xi^\rho_{i\mu}\xi^1 _{i\mu}M_\rho +\sqrt{\alpha N} \tilde{\xi}^\rho_{i\mu}\xi^1 _{i\mu}M_\rho)+\\
&\qquad+\nonumber \sum_{\rho=1}^K(\sqrt{\alpha N}\xi^\rho_{i\mu}\xi^1 _{i\mu}\tilde M_\rho+\alpha N \tilde{\xi}^\rho_{i\mu}\xi^1 _{i\mu}\tilde M_\rho)\Big]\\
&= (\mathcal {S} + \mathcal {N})/2N,
\label{eq:NS}
\end{align}
where in the first line of eq.~(\ref{eq:stability}) we used the trivial identity $(\xi^1_{i\mu})^2=1$ and in eq.~(\ref{eq:NS}) we split the energy contribution $h_{i\mu}\xi_{i \mu}^1$ into a signal $\mathcal {S}$ and a noise $\mathcal {N}$ term:
\begin{align}
\label{eq:S}
\mathcal {S} &= M_1,
\end{align}
\begin{align}\nonumber
\label{eq:N}
\mathcal {N} &= \sqrt{\alpha N}\Big[ \tilde{\xi}^1_{i\mu}\xi^1 _{i\mu}M_1+\sum_{\rho=2}^K ( \frac{ \xi^\rho_{i\mu}\xi^1 _{i\mu} M_\rho}{\sqrt{\alpha N}} +\tilde{\xi}^\rho_{i\mu}\xi^1 _{i\mu}M_\rho ) +\\
&\qquad+ \sum_{\rho=1}^K( \xi^\rho_{i\mu}\xi^1 _{i\mu}\tilde M_\rho+ \sqrt{\alpha N} \tilde{\xi}^\rho_{i\mu}\xi^1 _{i\mu}\tilde M_\rho) \Big].
\end{align}
We now compare the scaling behaviours of these two terms, by computing their ratio. 
We anticipate that this ratio depends on the realization of the noisy patterns and we should average in some way over the variables $\tilde{\xi}^\rho_{i\mu}$. If not interested in the magnitude of fluctuations (mirroring the statistical mechanical side, where the model is kept mean-field and analyzed at the replica symmetric level), one can simply consider the ratio $\mathbb E_{\tilde{\xi}}(\mathcal {S})/\mathbb E_{\tilde{\xi}}(\mathcal {N})$, where $\mathbb E_{\tilde{\xi}}(\cdot)$ is the average over the internal noise realizations $\tilde{\xi}^\rho_{i\mu}$. In this way, fluctuations are averaged out and we are only left with the magnitudes of the first moment.
Let us now turn to the evaluation of the scaling behaviours of $\mathcal {S}$ and $\mathcal {N}$ in \eqref{eq:S} and \eqref{eq:N}, respectively.
\newline
First, under the perfect retrieval hypothesis, we have $M_1=1$, whence
\begin{align}
\mathcal {S} =M_1=1.
\end{align}
As for $\mathcal {N}$, we can preliminary notice that, among its five contributions appearing in \eqref{eq:N}, the second term ${\sum_{\rho>1} \xi^\rho_{i\mu}\xi^1 _{i\mu}M_\rho}$ can be neglected as it is vanishing as $\mathcal  O(N^{-1/2})$ in the thermodynamic limit. This can be seen by expanding the magnetizations, that is
\begin{align}
\begin{split}
\sum_{\rho=2}^K &\xi^\rho_{i\mu}\xi^1 _{i\mu}M_\rho=\frac1{N^2}\sum_{\rho=2}^K \sum_{j,\nu=1}^{N,N}\xi^\rho_{i\mu}\xi^1_{i\mu}\xi^\rho_{j\nu}\xi^1_{j\nu}=\\&
=\frac1{N^2}\sum_{\rho=2}^K \Big(\xi^\rho_{i\mu}\xi^1_{i\mu}\xi^\rho_{i\mu}\xi^1_{i\mu}+\sum_{\substack{\nu=1 \\ \nu\neq\mu}}^N \xi^\rho_{i\mu}\xi^1_{i\mu}\xi^\rho_{i\nu}\xi^1_{i\nu}+\\
&\qquad\qquad\qquad\qquad+\sum_{\substack{\nu, j=1 \\ j \neq i}}^{N,N} \xi^\rho_{i\mu}\xi^1_{i\mu}\xi^\rho_{j\nu}\xi^1_{j\nu} \Big)\\&
=\frac1{N^2}\Big[\sum_{\rho=2}^K (\xi^\rho_{i\mu})^2(\xi^1_{i\mu})^2+\sum_{\rho=2}^K \sum_{\substack{\nu=1 \\ \nu\neq\mu}}^N \xi^\rho_{i\mu}\xi^1_{i\mu}\xi^\rho_{i\nu}\xi^1_{i\nu}+\\
&\qquad\qquad\qquad\qquad\sum_{\rho=2}^K \sum_{\substack{\nu, j=1 \\ j \neq i}}^{N,N} \xi^\rho_{i\mu}\xi^1_{i\mu}\xi^\rho_{j\nu}\xi^1_{j\nu} \Big],
\end{split}
\end{align}
and checking that the first term in square brackets is of order $\mathcal  O (N)$, due to the trivial equality $(\xi^\rho_{i\mu})^2=1$ and to the fact that the sum includes $K-1$ terms with $K\sim \alpha N$; the remaining two terms can be looked at as the displacement covered by simple random walkers performing, respectively, $\sim N^2$ and $\sim N^3$ steps on a linear chain, in such a way that for large enough $N$ they are Gaussian distributed with standard deviation of order $\mathcal  O (N)$ and $\mathcal  O (N^{3/2})$, respectively.
\newline
Therefore, in the large $N$ limit, the leading contribution in the noise term (\ref{eq:N}) is given by
\begin{align}\nonumber
\mathcal {N} \underset{N\to \infty}{\sim} &\sqrt{\alpha N}\tilde{\xi}^1_{i\mu}\xi^1_{i\mu}M_1+\sum_{\rho=2}^K \sqrt{\alpha N} \tilde{\xi}^\rho_{i\mu}\xi^1 _{i\mu}M_\rho+\\
&+\sum_{\rho=1}^K \sqrt{\alpha N} \xi^\rho_{i\mu}\xi^1 _{i\mu}\tilde M_\rho+\sum_{\rho=1}^K\alpha N \tilde{\xi}^\rho_{i\mu}\xi^1 _{i\mu}\tilde M_\rho,
\end{align}
and, when taking the average with respect to the pattern internal noise, only the last term survives, since it is the only one with even powers of $\tilde{\xi}^{\rho}_{i\mu}$.
\newline
Then, focusing on the last term, we get
\begin{align}
\begin{split}
\sum_{\rho=1}^K\alpha N \tilde{\xi}^\rho_{i\mu}\xi^1 _{i\mu}\tilde M_\rho&=\frac\alpha  N \sum_{\rho=1}^K \sum_{j,\nu=1}^{N,N} \tilde{\xi}^\rho_{i\mu}\xi^1 _{i\mu} \tilde{\xi}^\rho_{j\nu}\xi^1 _{j\nu},
\end{split}
\end{align}
and, introducing the variables $\vardbtilde{\xi}^\rho_{i\mu} =\xi^1_{i\mu}\tilde{\xi}^\rho_{i\mu}$, which are obviously Gaussian-distributed,
\begin{align}\label{eq:explicit}
\begin{split}
\sum_{\rho=1}^K&\alpha N \tilde{\xi}^\rho_{i\mu}\xi^1_{i\mu}\tilde M_\rho=\frac\alpha N \sum_{\rho=1}^K \sum_{j,\nu=1}^{N,N} \vardbtilde{\xi}^\rho_{i\mu} \vardbtilde{\xi}^\rho_{j\nu}=\\
&=\frac{\alpha}{N} \sum_{\rho=1}^K \left( \vardbtilde{\xi}^\rho_{i\mu}\vardbtilde{\xi}^\rho_{i\mu}+\sum_{\substack{\nu=1 \\ \nu\neq\mu}}^N \vardbtilde{\xi}^\rho_{i\mu} \vardbtilde{\xi}^\rho_{i\nu}+\sum_{\substack{\nu, j=1 \\ j\neq i}}^{N,N}   \vardbtilde{\xi}^\rho_{i\mu} \vardbtilde{\xi}^\rho_{j\nu} \right)=\\
&=\frac\alpha N \sum_{\rho=1}^K \left( \vardbtilde{\xi}^\rho_{i\mu}\right)^2+\frac\alpha N \sum_{\rho=1}^K \Big(\sum_{\substack{\nu=1 \\ \nu\neq\mu}}^N \vardbtilde{\xi}^{\rho}_{i\mu} \vardbtilde{\xi}^\rho_{i\nu}+\sum_{\substack{\nu, j=1 \\ j\neq i}}^{N,N} \vardbtilde{\xi}^\rho_{i\mu}\vardbtilde{\xi}^\rho_{j\nu}\Big).
\end{split}
\end{align}

Furthermore, in the expression \eqref{eq:explicit}, only the first term in the last line gives non-vanishing contribution (since the other two terms are product of uncorrelated random variables). Therefore
\begin{align}
\mathbb E_{\tilde{\xi}}(\mathcal {N}) =\frac\alpha N \sum_{\rho=1}^K \mathbb E_{\tilde{\xi}}(\vardbtilde{\xi}^\rho _{i\mu})^2\sim \alpha\frac KN=\alpha^2,
\end{align}
which is $\mathcal  O (1)$, that is the same scaling behaviour of the signal $M_1$.

The arguments just exposed allow us to introduce the ``pattern recognition power'' as the maximal extent of noise that can affect the information encoded by patterns (supposed $\mathcal {O}(1)$) still allowing pattern retrieval. This is strongly related to the memory storage: if we load the network with $K \sim N^3$ patterns then the pattern recognition power is $\mathcal  O(N^{0})$, if $K \sim N^2$, then the pattern recognition power is $\mathcal  O(N^{1/4})$, if $K \sim N^1$, then the pattern recognition power is $\mathcal  O(N^{1/2})$, and so on.
	Therefore, if $K \sim N^3$ the pattern recognition power of this net is the same as the one of the standard Hopfield model in high load, but if we sacrifice pattern storage letting $K \sim N^1$, then the pattern recognition power of this net is much higher than the one of the standard Hopfield model.

\section{Replica-Symmetric evaluation of the quenched pressure}
In this Appendix, we report the technical details underlying the solution of the model provided in the main text. Here, we will adopt a formal style to stress that the analysis is led by rigorous tools. In fact, beyond the signal-to-noise analysis performed in the previous Appendix, the numerical approach followed in the next Appendix and analytical non-rigorous methods based on the replica-trick (that we also carried out to check overall consistency, without presenting in details) that still retain a heuristic flavour, the problem can be actually addressed rigorously.

For completeness, let us recall the basic ingredients of the model.
\begin{Definition} The Hamiltonian function for the DAM neural network is
	\beq\label{Alemannation1}
	H_{\textrm{DAM}} = -\frac{1}{2N^3}\sum_{\rho=1}^K\left(\sum_{i, \mu=1}^{N,N} \eta^\rho_{i \mu} \sigma_i\tau_{\mu} \right)^2,
	\eeq
	where $\sigma_i,\tau_{\mu} \in \{ -1, +1 \}$ for $i, \mu=1,...,N$ and $K=\alpha N$.
\end{Definition}
Following the preliminary analysis by the SNR (see Appendix B), the noisy tensors $\boldsymbol{\eta}$ is given by
\begin{Definition} The interaction strength for the dimer $(\sigma_i,\tau_{\mu})$ is defined as
	\beq \label{eq:prescription}
	\eta^\rho_{i \mu}=  \,\xi_{ij}^{\rho}  +  \sqrt{K} \tilde{\xi}^\rho_{i \mu} =  \,\xi^\rho_i  \xi^\rho_\mu +  \sqrt{K} \tilde{\xi}^\rho_{i \mu},
	\eeq
	where the $\xi^{\rho}_{i}$'s are i.i.d. drawn from $\mathbb{P}(\xi^\rho_{i}=\pm1)=1/2$, while the $\tilde{\xi}^\rho_{i \mu}$'s are i.i.d. drawn from $\mathbb{P}(\tilde{\xi}^\rho_{i \mu}) =  \mathcal {N} (0,1)$.
\end{Definition}

In this Appendix, we extend Guerra's interpolating scheme \cite{Barra-JSP2010} to deal with these dense networks. This technique works directly on the main quantity of interest in the Statistical Mechanical analysis, namely the quenched pressure associated to the cost function (\ref{Alemannation1}), as introduced in the next
\begin{Definition}
	The quenched pressure density associated to the Hamiltonian (\ref{Alemannation1}) is defined as
	\beq
	A(\alpha,\beta) \equiv \lim_{N\to \infty}\frac{1}{N}\bbE_{\boldsymbol{\eta}} \log Z,
	\eeq
	where $Z$ is the partition function associated to the Hamiltonian (\ref{Alemannation1}) given by
	\beq
	Z \equiv \sum_{\boldsymbol{\sigma},\boldsymbol{\tau}}\exp \left[ \frac{\beta}{2N^3}\sum_{\rho=1}^K\left(\sum_{i,\mu=1}^{N,N} \eta^\rho_{i \mu} \sigma_i \tau_\mu \right)^2 \right],
	\eeq
	and $\mathbb E_{\boldsymbol{\eta}}$ denotes the quenched average over the realizations of the tensor $\boldsymbol{\eta}$: for a generic function $f$ of the tensor elements $\{ \eta_{i\mu}^\rho \}$ this average is defined as
	\beq
	&\bbE_{\boldsymbol{\eta}} f( \{ \eta_{i\mu}^\rho \} ) =\frac{1}{2^N} \prod_{\rho=1}^K \sum_{\xi^\rho_1\cdots\xi^\rho_N=\pm1} \int \left(\prod_{i,\mu=1}^{N,N} D\tilde{\xi}^\rho_{i \mu} \right) f( \{ \eta_{i\mu}^\rho \} )\\ \nonumber&
	\textrm{with}\\\nonumber
	&\qquad Dx \equiv \frac{dx}{\sqrt{2\pi}}e^{-x^2/2}.
	\eeq
\end{Definition}
We stress that the partition function can be written in terms of the magnetizations \eqref{eq:magnetizations} as
\begin{align}
Z= \sum_{\boldsymbol{\sigma},\boldsymbol{\tau}}\exp\Big[\frac{\beta N}2 \sum_{\rho=1}^K \left(M_\rho+\sqrt{K} \tilde M_\rho \right)^2\Big].
\end{align}
Since we are looking for the retrieval regime, we shall assume that only a single information pattern (say $\xi^1$) is candidate for the condensation. Then,
\begin{align}\label{eq:forinterp}\nonumber
Z=\sum_{\boldsymbol{\sigma},\boldsymbol{\tau}}\exp\Big[\frac{\beta N}2 (M_1+\sqrt{K} \tilde M_1)^2+\\\qquad+\frac{\beta N}2 \sum_{\rho=2}^K(M_\rho+\sqrt{K} \tilde M_\rho )^2\Big].
\end{align}
The magnetization $M_1$ associated to the retrieved pattern is of order $\mathcal  O(1)$, while all the others $M_\rho$ are $\mathcal  O(N^{-1})$. This implies that, in the thermodynamic limit, we can neglect the subleading contributions, so that (note that this decomposition leads to the same results also in case of $M_1\sim \mathcal  O (N^{-1})$, i.e. when the network fails to retrieve the presented pattern. In that case, the overlap of the network with the external noise source $\tilde{\xi}^1$ is not negligible w.r.t. to the signal part. However, discarding it (i.e. only taking the last sum in \eqref{eq:forinterp}) would lead to a negligible error in the thermodynamic limit. Indeed, it is straightforward to check that the former is of order $\mathcal  O(1)$, while the remaining contributions scale as $\mathcal  O(N)$)
\begin{align}\nonumber
\label{eq:partition_touse}
Z&\underset{N\to \infty}{\sim} \sum_{\boldsymbol{\sigma},\boldsymbol{\tau}}\exp\left(\frac{\beta N}2 M_1^2+\frac{\alpha\beta N^2}2 \sum_{\rho=2}^K  \tilde M_\rho^2\right)=\\&=
\sum_{\boldsymbol{\sigma},\boldsymbol{\tau}}\int D\textbf{z}\exp\Big(\frac{\beta N}2 m_\sigma^2 m_\tau^2+\nonumber\\
&\qquad\qquad+\sqrt{ \alpha}\frac{ \beta}{N}\sum_{\rho = 2}^K\sum_{i,\mu=1}^{N,N} \tilde{\xi}^\rho_{i \mu} \sigma_i\tau_\mu z_\rho \Big),
\end{align}
where, in the second line, we used the Hubbard-Stratonovich linearization (by this, $D\textbf{z}$ is the ($K-1$)-dimension $\mathcal  N(0,\beta^{-1})$ Gaussian measure) and we posed $M_1=m_\sigma m_\tau$ with
\bes\label{eq:mattisfact}
m_\sigma&\equiv \frac{1}{N}\sum_{i=1}^N \xi^1_{i} \sigma_i,\  \ \ \ m_\tau\equiv \frac{1}{N}\sum_{\mu=1}^N \xi^1_{\mu} \tau_\mu,
\ees
reflecting the factorization of the signal part of the interaction strength, i.e. $\xi^1_{i\mu}=\xi^1_i \xi^1_\mu$. The expression in the last line of \eqref{eq:partition_touse} is the starting point for our interpolation procedure.
\begin{Definition}
	Guerra's interpolating function related to the quenched pressure of the DAM cost function (\ref{Alemannation1}) is
	\beq
	\nonumber
	\mathcal  A(t)=\frac{1}{N}\bbE_{\boldsymbol{\eta}} \log Z_t,
	\eeq
	where
	\begin{align}\nonumber
	\label{eq:Z_t}
	Z_t& \equiv  \sum_{\boldsymbol{\sigma},\boldsymbol{\tau}}\int D \textbf{z}\exp\Big(t\frac{\beta}{2} Nm^2_\sigma m^2_\tau +\\\nonumber
	&\qquad\qquad+ \sqrt{t}\sqrt{ \alpha }\frac{\beta}{N}\sum_{\rho = 2}^K\sum_{i,\mu=1}^{N,N} \tilde{\xi}^\rho_{i \mu} \sigma_i \tau_\mu z_\rho +\\
	&\qquad\qquad+ \sqrt{1-t} ~\mathcal W + (1-t)~\mathcal D\Big)
	\end{align}
	is the generalized partition function and $\mathcal W$ and $\mathcal D$ are defined as
	\bes
	\mathcal D \equiv& N C_1 m_\sigma +  N C_2 m_\tau+ C_6\sum_{\rho = 2}^K \frac{z^2_\rho}{2},\\
	\mathcal W \equiv& C_3\sum_{i=1}^N \tilde{\xi}^{(1)}_i\sigma_i+ C_4\sum_{\mu=1}^N \tilde{\xi}^{(2)}_\mu \tau_\mu + C_5\sum_{\rho = 2}^K \tilde{\xi}_\rho z_\rho,
	\ees
	where the external fields $\tilde{\xi}^{(1)}_i$, $\tilde{\xi}^{(2)}_\mu$ and $\tilde{\xi}_\rho$ are i.i.d. variables and $C_1,...,C_6$ are suitable constants to be set a posteriori.
\end{Definition}
\begin{Definition}\label{def:5}
		Given a generic function $F(\boldsymbol \sigma, \boldsymbol \tau, \boldsymbol z)$ of the neurons, the (generalized) Boltzmann average $\omega_t(F)$ is defined as
		\begin{align}\nonumber
		\omega_t (F) &\equiv Z_t^{-1}\sum_{\boldsymbol{\sigma},\boldsymbol{\tau}}\int D \textbf{z}\,F(\boldsymbol \sigma, \boldsymbol \tau, \boldsymbol z)\exp\Big(t\frac{\beta}{2} Nm^2_\sigma m^2_\tau\Big)\times\\\nonumber
		&\times\exp\Big(\sqrt{t}\sqrt{ \alpha }\frac{\beta}{N}\sum_{\rho = 2}^K\sum_{i,\mu=1}^{N,N} \tilde{\xi}^\rho_{i \mu} \sigma_i \tau_\mu z_\rho\Big)\times\\
		&\times\exp\Big(\sqrt{1-t} ~\mathcal W + (1-t)~\mathcal D\Big).
		\end{align}
	\end{Definition}

Note that, as standard in Guerra's interpolation techniques (see \cite{Barra2006} for ferromagnets, \cite{Guerra} for spin glasses and \cite{Barra-JSP2010} for neural networks), in the function $\mathcal A(t)$, the interpolating parameter appears with different exponents ($1$ and $1/2$) mirroring the nature of the interaction (ferromagnetic and glassy, respectively) and, ultimately, the need to apply Wick's theorem.
\begin{Remark}
	Guerra's interpolating function evaluated at $t=1$ corresponds to the original quenched pressure, in such a way that its explicit expression can be recovered via a simple sum rule by using the Fundamental Theorem of Calculus, i.e.
	\beq\nonumber
	A(\alpha,\beta)=&\lim_{N\to\infty}\mathcal A(t=1)=\\
	=&\lim_{N\to\infty}\left(\mathcal A(t=0)+\int_0^1 dt\, \partial_t\mathcal A(t)\right).
	\label{eq:sumrule}
	\eeq
\end{Remark}
Therefore we  now have to evaluate $\partial_t\mathcal A$ and $\mathcal A(0)$. This calculation is rather lengthy and goes along the same line as \cite{Barra-JSP2010} without requiring any particular operation; for this reason, we shall report the explicit passages related only to the second term (that is, the most complex) in the argument of the exponential in $\mathcal A(t)$ and just for illustrative purposes. Then, let us pose
	\beq\nonumber
	\mathcal A^{\textrm{(2)}}(t) &\equiv \frac{1}{N}\bbE_{\boldsymbol{\eta}} \log \sum_{\boldsymbol{\sigma},\boldsymbol{\tau}}\int D \textbf{z}\times\\
	&\times\exp\left(\sqrt{t}\sqrt{ \alpha }\frac{\beta}{N}\sum_{\rho = 2}^K\sum_{i,\mu=1}^{N,N} \tilde{\xi}^\rho_{i \mu} \sigma_i \tau_\mu z_\rho\right),
	\eeq
	and define the generalized Boltzmann average $\omega_t ^{(2)}$ as in Def. \ref{def:5} (of course, when dealing with the generalized pressure $\mathcal  A(t)$ rather than $\mathcal  A^{(2)}(t)$, we have to replace $\omega^{(2)}_t$ with the corresponding Boltzmann average $\omega_t$).
		Thus, deriving with respect to $t$, we get
	\begin{align}\nonumber
	\partial_t& \mathcal A^{\textrm{(2)}}(t)=\frac{\sqrt{\alpha}}{2 \sqrt{t}} \frac{\beta}{N^2} \sum_{\rho =2}^{K} \sum_{i, \mu=1}^{N,N}  \mathbb{E}_{\boldsymbol{\eta}} ~\tilde{\xi}^\rho_{i \mu} ~ \omega_t ^{(2)}(\sigma_i \tau_{\mu} z_{\rho})\\\nonumber
	&=\frac{\alpha \beta^2 }{2 N^3} \sum_{\rho =2}^{K} \sum_{i, \mu=1}^{N,N}  \mathbb{E}_{\boldsymbol{\eta}}\left[ \omega_t ^{(2)}(\sigma_i^2 \tau_{\mu}^2 z_{\rho}^2) - \omega_t ^{(2)}(\sigma_i \tau_{\mu} z_{\rho})  ^2   \right]\\\nonumber
	&=\frac{\alpha \beta^2 }{2 N^3} \sum_{\rho =2}^{K} \sum_{i, \mu=1}^{N,N}  \mathbb{E}_{\boldsymbol{\eta}} \omega_t ^{(2)}(z_{\rho}^2) - \frac{\alpha^2 \beta^2 }{2} \langle q_{12} r_{12} p_{12} \rangle  _t ^{(2)}   \\\label{eq:new_expl}
	&=\frac{\alpha^2 \beta^2 }{2} \langle p_{11} \rangle _t ^{(2)}- \frac{\alpha^2 \beta^2 }{2} \langle q_{12} r_{12} p_{12} \rangle_t^{(2)}.
	\end{align}
	Here, in the second passage we applied Wick's theorem, in the third passage we exploited the Boolean nature of the $\sigma$ and $\tau$ variables, we defined $\langle \cdot \rangle_t ^{(2)} \equiv \mathbb{E}_{\boldsymbol \eta} \omega_t^{(2)} (\cdot)$ and introduced two-replica overlaps (one for each layer) to account for the quenched noise. More precisely:
\bes
q_{12}&\equiv \frac{1}{N}\sum_{i=1}^N \sigma^1_i\sigma_i^2,\\
r_{12}& \equiv \frac{1}{N}\sum_{\mu=1}^N \tau^1_\mu \tau_\mu^2,\\
p_{12}& \equiv \frac{1}{K-1}\sum_{\rho = 2}^K z^1_\rho z_\rho^2.
\ees
Further, the first term in (\ref{eq:new_expl}) can be cancelled by suitably choosing the constant $C_6$ (dedicated to tune the variance $z^2$).
	Repeating analogous calculations for the remaining terms making up $\mathcal A(t)$, overall we get
\bes
\label{eq:AA}
\partial_t\mathcal A&=\bbE_{\boldsymbol{\xi}} \Big[\frac{\beta}{2}\mv{m_\sigma^2m_\tau^2}_t + \frac{ \alpha^2\beta^2}2 \mv{p_{11}}_t+\\
&-\frac{ \alpha^2\beta^2}2 \mv{p_{12}q_{12}r_{12}}_t-C_1 \mv{m_\sigma}_t-C_2 \mv{m_\tau}_t+\\
&-\frac{\alpha C_6}2 \mv{p_{11}}_t-\frac{C_3^2}2 - \frac{C_4^2}2 +\frac{C_3^2}2\mv{q_{12}}_t+\\
&+\frac{C_4^2}2\mv{r_{12}}_t-\frac{\alpha C_5^2}2\mv{p_{11}}_t+\frac{\alpha C_5^2}2\mv{p_{12}}_t\Big],
\ees
where now $\langle \cdot \rangle_t  \equiv \mathbb{E}_{\boldsymbol \eta} \omega_t(\cdot)$ and the quenched average $\bbE_{\boldsymbol{\xi}}$ applies only on the Boolean variables $\boldsymbol{\xi}$ as the Gaussian variables $\tilde{\boldsymbol{\xi}}$ have been already averaged out (via Wick's theorem).
As anticipated, a trivial simplification can be implemented by setting ${ C_6}={\alpha\beta^2}-{ C_5^2}$ in such a way that we get rid of $\mv{p_{11}}$.
A further simplification can be obtained asking for vanishing fluctuations for the order parameters, as prescribed by the RS approximation in the thermodynamic limit. Then, calling $(\overline m_\sigma,\ \overline m_\tau)$ and $(q,\ p,\ r)$ the RS values for, respectively, the Mattis magnetizations and the overlaps, the corresponding probability distributions in the thermodynamic limit satisfy
\bes
&\lim_{N\to\infty} \mathbb P(m_{\sigma})=\delta(m_{\sigma}-\overline m_\sigma),\\
&\lim_{N\to\infty} \mathbb P(m_{\sigma})=\delta(m_{\tau}-\overline m_\tau),\\
&\lim_{N\to\infty} \mathbb P(q_{12})=\delta(q_{12}-q),\\
&\lim_{N\to\infty} \mathbb P(p_{12})=\delta(p_{12}-p),\\
&\lim_{N\to\infty} \mathbb P(r_{12})=\delta(r_{12}-r).
\ees
Denoting with $\Delta$ the fluctuation of the generic observable w.r.t. its thermodynamic value (e.g., $\Delta = m_{\sigma} - \overline{m}_\sigma$), we can recast the interaction terms appearing in \eqref{eq:AA} as
%
\bes
\mv{p_{12}q_{12}r_{12}} _t&= -2pqr +pq\mv{r_{12}}_t+pr\mv{q_{12}}_t+\\
&\qquad\qquad\qquad\qquad\qquad+rq\mv{p_{12}}_t+\mathcal {O}(\Delta^2),\\
\mv{m_\sigma^2m_\tau^2}_t&=-3\overline m_\sigma^2\overline m_\tau^2 +2 \overline m_\sigma^2\overline m_\tau \mv{m_\tau}_t+\\
&\qquad\qquad\qquad\qquad+2 \overline m_\tau^2\overline m_\sigma \mv{m_\sigma}_t+\mathcal {O}(\Delta^2).
\ees
Moreover, since in the RS regime fluctuations vanish, we can disregard terms $\mathcal  O(\Delta^2)$, obtaining
\bes
\mv{p_{12}q_{12}r_{12}}_t &= -2pqr +pq\mv{r_{12}}_t+pr\mv{q_{12}}_t+rq\mv{p_{12}}_t,\\
\mv{m_\sigma^2m_\tau^2}_t&=-3\overline m_\sigma^2\overline m_\tau^2 +2 \overline m_\sigma^2\overline m_\tau \mv{m_\tau}_t+2 \overline m_\tau^2\overline m_\sigma \mv{m_\sigma}_t.
\ees
Replacing these expressions inside the streaming term, and choosing our free parameters as
\bes
\label{eq:Cis}
C_1&=\beta \overline m_\sigma \overline m_\tau^2,\\
C_2&=\beta \overline m_\tau \overline m_\sigma^2,\\
C_3^2&=\alpha^2\beta ^2 pr,\\
C_4^2&=\alpha^2 \beta^2 pq,\\
C_5^2&=\alpha \beta^2 qr,\\
C_6&=\alpha\beta^2(1-qr),\\
\ees
we reach the simple result
\beq
\partial_t\mathcal A= \frac{\alpha^2 \beta^2}{2} p(2qr -q-r)-\frac{3}{2}\beta \overline m_\sigma^2 \overline m_\tau^2,
\eeq
which is independent on $t$, so that the integration is trivial. Now, we must evaluate the one-body term:
\bes
\mathcal A(0)&=\frac{1}{N}\bbE_{\boldsymbol\eta} \log \sum_{\boldsymbol{\sigma}, \boldsymbol{\tau}}\int D \textbf{z}\times\\
&\times\exp \Big(N C_1 m_\sigma +  N C_2 m_\tau+ C_6\sum_{\rho=1}^K \frac{z^2_\rho}{2}\Big)\times\\
&\times\exp\Big(C_3\sum_{i=1}^N \tilde{\xi}_i^{(1)} \sigma_i+ C_4\sum_{\mu=1}^N \tilde{\xi}_\mu^{(2)} \tau_\mu + C_5\sum_{\rho=1}^K \tilde{\xi}_\rho z_\rho \Big).
\ees
With straightforward computations and recalling the choices \eqref{eq:Cis} for the $C_i$ coefficients, we have
\bes
\mathcal A(0)=&2\log2 +\bbE_x \log\cosh\big[\alpha \beta x \sqrt{p r}+ \beta \overline m_\sigma \overline m_\tau^2 \big]+\\
&+\bbE_x \log\cosh\big[\alpha\beta x \sqrt{ p q}+ \beta \overline m^2_\sigma \overline m_\tau \big]+\\
&+\frac{\alpha^2\beta}{2} \frac{ q r}{1-\alpha\beta(1-qr)}-\frac{\alpha}{2} \log [ 1-\alpha\beta(1-qr)],
\ees
where $x$ is a standard Gaussian variable.
Exploiting the sum rule \eqref{eq:sumrule} 
we have  the final result
\begin{Theorem}
	In the thermodynamic limit, under the replica symmetric approximation, the quenched pressure density related to the cost function (\ref{Alemannation1}) can be expressed in terms of the natural order parameters of the model (i.e. the two Mattis magnetizations for the visible and mirror layers and the three two-replica overlaps of the visible, hidden and mirror layers) as follows
	\bes
	A_{\RS}=&2\log2 +\bbE_x \log\cosh\big[\alpha \beta x \sqrt{p r}+ \beta \overline m_\sigma \overline m_\tau^2 \big]+\\
	&+\bbE_x \log\cosh\big[\alpha\beta x \sqrt{ p q}+ \beta \overline m^2_\sigma \overline m_\tau \big]+\\
	&+\frac{\alpha^2\beta}{2} \frac{ q r}{1-\alpha\beta(1-qr)}-\frac{\alpha}{2} \log[1-\alpha\beta(1-qr)]+\\
	&+\frac{\alpha ^2\beta^2}{2} p(2qr -q-r)-\frac{3}{2}\beta \overline m_\sigma^2 \overline m_\tau^2.
	\ees
	Its extremization selects the maximum entropy solutions that minimize the cost function \ref{Alemannation1} and yields to the self-consistent equations (11)-(15).
\end{Theorem}

As a final remark, we note that, from a machine-learning perspective, beyond signal detection (involving the Mattis magnetizations), also quenched noise is to be estimated and, since the latter is carried by the overlaps, a first estimate can be obtained by a Plefka-like expansion of the free energy in the high (fast)-noise limit (see e.g., \cite{B00,JR,SM} and reference therein).\\

\section{Replica symmetric phase diagram}

\begin{figure}[tb]
	\begin{center}
		\includegraphics[scale=0.35]{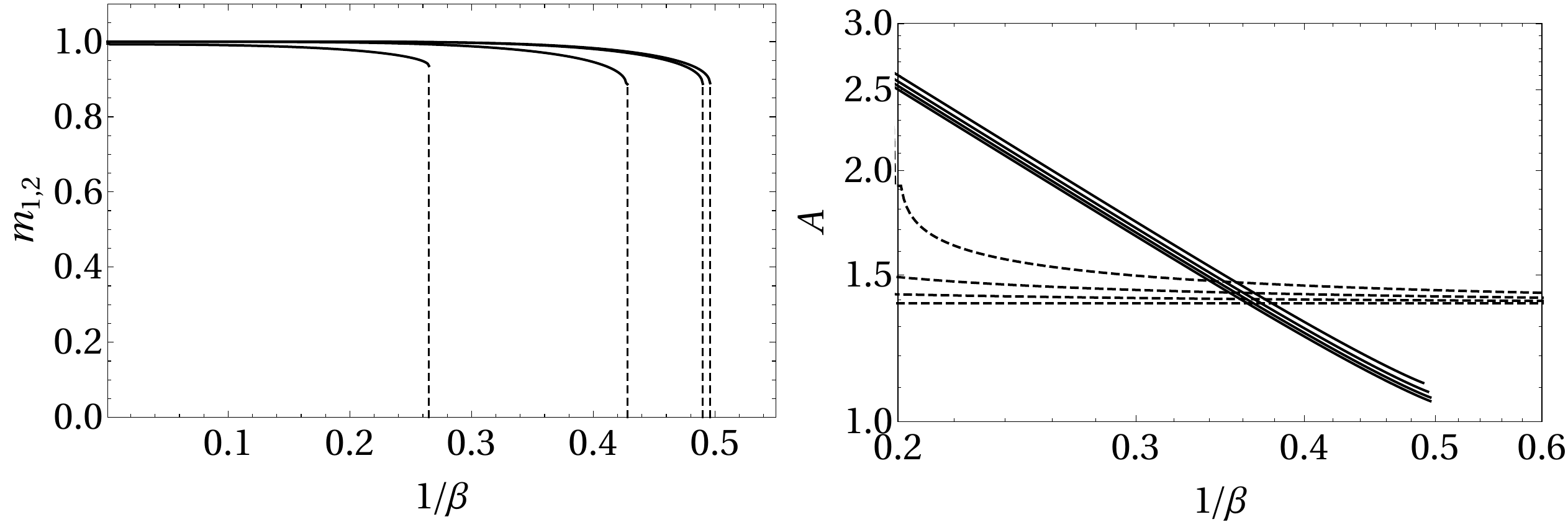}
		\caption{Mattis magnetization(s) and free-energy. Left: the plot shows the Mattis magnetization $m$ (we stress that, on the self-consistency solutions, $m_1=m_2$) as a function of the fast noise $1/\beta$ for various storage capacity values ($\alpha = 0, 0.20, 0.40, 0.50$, going from the right to the left). The vertical dotted lines indicates the jump discontinuity identifying the critical noise level $1/\beta(\alpha)$ that traces the boundary between the retrieval region and the pure spin-glass phase. Right: the plot shows the corresponding pressure as a function of the fast noise level $1/\beta$ at the storage capacity values $\alpha = 0, 0.10, 0.15, 0.20$ (going from the bottom to the top) in the retrieval (continuous lines) and spin-glass (dotted lines) states.
				Note that the sampled $\alpha$-values are different among the two plots for a matter of best visualization (for too low values of $\alpha$ all the magnetizations heavily overlap and it is hard to distinguish them by eye inspection). Note: the solutions always share the symmetry $m_1=m_2$.}
		\label{Fig:PD}
	\end{center}
\end{figure}

The phase diagram shown in Fig.~2 of the main text exhibits four qualitatively different phases as explained hereafter:
	\begin{itemize}
		\item{Ergodic phase (E)}\\
		The ``fast'' noise $1/\beta$ in the system is too strong for the neurons to reciprocally feel
		each other, in such a way that they tend to behave randomly and no
		emergent collective property can be appreciated. In this region, the solution of the self-consistency equations [i.e., eqs. (11)-(15) in the main text] which maximizes the pressure [i.e., eq. (10) in the main text] is given by $m = 0, ~ q = 0$.
		\item{Spin-glass phase (SG)}\\
		The ``slow'' noise $\alpha$ is too large for the neurons to correctly handle the whole set of patterns, and again the system fails to retrieve information, although the thermalized configurations are not purely random. In this region, the solution of the self-consistency equations which maximizes the pressure is given by $m = 0, ~  q \neq 0$.
		\item{Retrieval phase (R1)}\\
		Both ``fast'' and ``slow'' noise are small enough for neural collective capabilities to spontaneously appear. In particular, the most likely configurations, namely those corresponding to the global maxima of the pressure, are those corresponding to stored patterns. In this region the solution of the self-consistent equations which maximizes the pressure is therefore given by $m \neq 0, ~ q \neq 0$.
		\item{Retrieval phase (R2)}\\
		Both ``fast'' and ``slow'' noise are still relatively small hence neural collective capabilities can still spontaneously appear. However, here configurations corresponding to stored patterns are only local maxima of the pressure in such a way that patterns can be retrieved as far as the initialization of the system is not too far (in the sense of the Hamming distance) with respect to the target pattern. In this region the self-consistent equations admit as solution $m \neq 0, ~ q \neq 0$
		as well as $m = 0, q \neq 0$, both corresponding to maxima of the pressure, the former being local maxima, the latter being global ones.
	\end{itemize}

%

A sketch of the analysis underlying the definition of the various regions is provided in Fig. \ref{Fig:PD}.

\begin{figure}[bt]
	\begin{center}
		\includegraphics[scale=0.26]{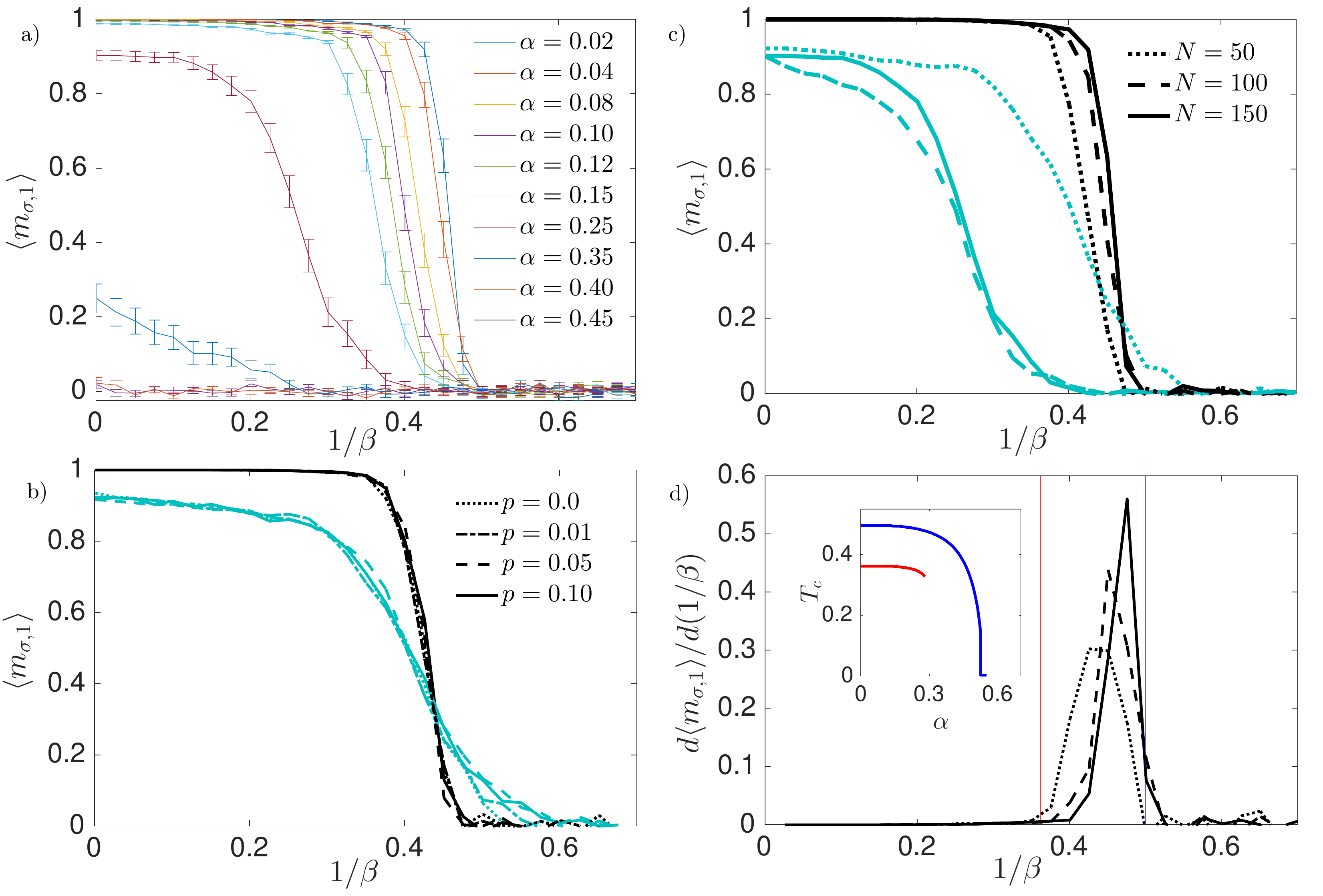}
		\caption{This figure shows results obtained through Monte Carlo simulations. Seeking for clarity, only $\langle m_{\sigma, \rho} \rangle$ is shown, but quantitatively analogous values are obtained also for $\langle m_{\tau, \rho} \rangle$. Errorbars (reported only in panel $a$, seeking for clarity) stem from the average of thermal noise and quenched noise. All cases depicted here correspond to $Q = 100$ realizations. Panel a: Expected Mattis magnetization $\langle m_{\sigma, \rho} \rangle$ for $N=150$ and $p=0.01$ as a function of $1/\beta$ and for different values of $\alpha$ as explained by the legend. Panel b: Comparison of the expected Mattis magnetization $\langle m_{\sigma, \rho} \rangle$ for different initial configurations $p=0.01, 0.05, 0.1$ (depicted in different line style as explained by the legend), for fixed $N=150$ and for $\alpha=0.02$ (dark curves) and $\alpha=0.25$ (bright curves). Panel $c$: Comparison of the expected Mattis magnetization $\langle m_{\sigma, \rho} \rangle$ for different sizes $N=50, 100, 150$ (depicted in different line style as explained by the legend), for fixed $p=0.01$ and for $\alpha=0.02$ (dark curves) and $\alpha=0.25$ (bright curves). Panel $d$: The critical noise $T_c$ is estimated by taking the discrete derivative of the expected Mattis magnetization $\langle m_{\sigma, \rho} \rangle$ with respect to the noise and by selecting the value of noise $1/\beta$ (if any) where the derivative peaks. Such estimates are obtained for $N=50, 100, 150$ (same legend as panel $c$) and for $p=0.01$, $\alpha =0.02$; the corresponding theoretical values are recalled in the inset.}
		\label{Fig:MC}
	\end{center}
\end{figure}

\section{Monte Carlo simulations}
We performed Monte Carlo simulations to mimic the evolution of a finite-size DAM network made of $N$ neurons interacting ($P=4$)-wisely and $K=\alpha N$ patterns, described by the cost function (\ref{Alemannation1}).
We first fixed the parameters $(N, \alpha, \beta)$ where\ $K$ has to be meant as the integer part of $ \alpha N$.
Then, we drew the i.i.d. Boolean variables $\xi_i^{\rho}$, with $i=1,...,N$ and $\rho=1,...,K$ as well as the related Gaussian noise $\tilde{\xi}_{i \mu}^{\rho}$ with $i, \mu=1,...,N$ and $\rho=1,...,K$. Then, the tensor $\boldsymbol{\eta}$ 
is built following the prescription (\ref{eq:prescription}).
Next, we initialize the system configuration in such a way that $\boldsymbol \sigma$ and $\boldsymbol \tau$ are aligned with $\boldsymbol \xi^1$, except for a fraction $p$ of misaligned entries, and we let the system evolve by a single spin-flip Glauber dynamics. Once the equilibrium state is reached, we collect data for the instantaneous Mattis magnetizations $m_{\sigma, \rho}=\sum_i \xi_i^{\rho} \sigma_i/N$ and $m_{\tau, \rho}=\sum_\mu \xi_\mu^{\rho} \tau_i/N$ 
to obtain the thermal average referred to as $\langle {m}_{\sigma,\rho} \rangle$ and $\langle{m}_{\tau,\rho} \rangle$, with $\rho=1,...,K$ (notice that, initially, one has $m_{\sigma, 1} = m_{\tau, 1} = 1 -2p$, while $m_{\sigma, \rho \neq 1}, m_{\tau, \rho \neq 1} \approx 0$). This is repeated for $Q=100$ different realizations of the patterns $\boldsymbol{\xi}$ and the noise $\tilde{\boldsymbol{\xi}}$, over which thermal averages are accordingly averaged. The resulting values provide our numerical estimate for the expectation of the Mattis magnetizations $\langle m_{\sigma, \rho} \rangle$ and $\langle m_{\tau, \rho}  \rangle$ to be compared with the solution of the self-consistent equations (14) and (15). Different parameters $(N, \alpha, \beta, p)$ are considered and, for each choice, the same procedure applies. A sample of our results for $N=150, p=0.01$ and different values of $\alpha, \beta$ is shown in Fig.\ref{Fig:MC}a, where one can check that the Mattis magnetization $m_{\sigma,1}$ corresponding to the retrieved pattern $\boldsymbol{\xi}^1$ vanishes at large values of the noise$T\equiv 1/\beta$ and/or at large values of $\alpha$; as expected from the theoretical analysis, the larger $\alpha$ and the smaller the critical temperature $T_c$ above which no retrieval takes place. We also notice that for $T=0$, when $\alpha$ increases beyond $\alpha \approx 0.3$ the magnetization abruptly vanishes. Then, in Fig.\ref{Fig:MC}b we compare results stemming from different choices of $p$ and for
$\alpha=0.02$ (dark curves) and $\alpha =0.25$ (bright curves): the initial configuration (as long as close enough to $\boldsymbol{\xi}^1$) does not influence quantitatively the final outcome. Next, in Fig.~\ref{Fig:MC}c we perform a finite-size-scaling considering, again, $\alpha=0.02$ (dark curves) and $\alpha =0.25$ (bright curves): the curves for $N=50$, $N=100$ and $N=150$ are slightly shifted and the shift gets more significant as $\alpha$ is increased. Finally, in Fig.~\ref{Fig:MC}d, main plot, we show the numerical derivative of $m_{\sigma,1}$ for $\alpha=0.02$ and for the three sizes analyzed before: the peak in the derivative can be used to estimate $T_c$ and this can in turn be compared with the theoretical results highlighted by the vertical lines and recalled in the inset.

\vspace{0.8cm}
The authors are grateful to Unisalento and Sapienza University of Rome (RG11715C7CC31E3D) for partial fundings.

\end{document}